**Title:** A General Temperature-Guided Language Model to Design Proteins of Enhanced Stability and Activity

**Short title:** Temperature-Aware Model for Protein Engineering

## Authors:

Fan Jiang[1†], Mingchen Li[2,3†], Jiajun Dong[4,5†], Yuanxi Yu[1†], Xinyu Sun[6,7†], Banghao Wu[1,8†], Jin Huang[1,8], Liqi Kang[1], Yufeng Pei[7], Liang Zhang[1], Shaojie Wang[4], Wenxue Xu[4], Jingyao Xin[4], Wanli Ouyang[2], Guisheng Fan[3], Lirong Zheng[1], Yang Tan[2,3], Zhiqiang Hu[9], Yi Xiong[8], Yan Feng[8], Guangyu Yang[8,10,11], Qian Liu[8], Jie Song[7*], Jia Liu[4*], Liang Hong[1,2,12*], Pan Tan[1,2*]

1. School of Physics and Astronomy, & Shanghai National Center for Applied Mathematics (SJTU Center), & Institute of Natural Sciences, Shanghai Jiao Tong University, Shanghai 200240, China
2. Shanghai Artificial Intelligence Laboratory, Shanghai 200030, China
3. School of Information Science and Engineering, East China University of Science and Technology, Shanghai 200240, China
4. Shanghai Institute for Advanced Immunochemical Studies and School of Life Sciences and Technology, ShanghaiTech University, Shanghai 201210, China
5. Guangzhou National Laboratory, No. 9 XingDaoHuanBei Road, Guangzhou International Bio Island, Guangzhou, Guangdong 510005, China
6. Department of Chemistry, University of Science and Technology of China, Hefei, Anhui 230001, China
7. Hangzhou Institute of Medicine, Chinese Academy of Sciences, Hangzhou, Zhejiang 310018, China
8. School of Life Sciences and Biotechnology, & State Key Laboratory of Microbial Metabolism, & Joint International Research Laboratory of Metabolic, Shanghai Jiao Tong University, Shanghai 200240, China
9. SenseTime Research, Shanghai 200233, China
10. Institute of Key Biological Raw Material, Shanghai Academy of Experimental Medicine, Shanghai 201401, China
11. Hzymes Biotechnology Co. Ltd, Wuhan, Hubei 430075, China
12. Zhanjiang Institute for Advanced Study, Shanghai Jiao Tong University, Shanghai 200240, China

[†] These authors contributed equally to this work.

[*] Corresponding author.

## Abstract

Designing protein mutants with both high stability and activity is a critical yet challenging task in protein engineering. Here, we introduce PRIME, a deep learning model, which can suggest protein mutants with improved stability and activity without any prior experimental mutagenesis data for the specified protein. Leveraging temperature-aware language modeling, PRIME demonstrated superior predictive ability compared to current state-of-the-art models on the public mutagenesis dataset across 283 protein assays. Furthermore, we validated PRIME's predictions on five proteins, examining the impact of the top 30-45 single-site mutations on various protein properties, including thermal stability, antigen-antibody binding affinity, and the ability to polymerize non-natural nucleic acid or resilience to extreme alkaline conditions. Remarkably, over 30% of the AI-recommended mutants exhibited superior performance compared to their pre-mutation counterparts across all proteins and desired properties. Moreover, we developed an efficient and effective method based on PRIME to rapidly obtain multi-site mutants with enhanced activity and stability. Hence, PRIME demonstrates broad applicability in protein engineering.

# Teaser:

PRIME model suggests superior protein mutants without prior data, achieving a success rate over 30% in key property modifications.

# Introduction

Proteins are fundamental constituents of living systems, playing crucial roles in a vast array of biological processes, spanning from enzyme catalysis (*1*) and cellular metabolism (*2*) to immune responses (*3*), signal transduction (*4*), and transport (*5*), among others. Beyond their biological significance, proteins are critical to numerous industries. In biomedicine, they serve as therapeutic agents and targets; in the food industry, they play roles in food processing and preservation; in brewing, they are essential to the production process; and in chemical engineering, they act as key catalysts for various reactions. Additionally, proteins are the cornerstone of in vitro diagnostic (IVD) tests, instrumental in the detection and monitoring of numerous diseases. However, proteins extracted from biological organisms, known as "wild type", often require modifications to make them suitable for industrial applications. This is primarily because the physicochemical conditions (e.g., temperature) in which these proteins need to function in industrial settings are often drastically different from their native biological contexts (*6, 7*). Therefore, to meet the demands of these diverse application scenarios, the proteins need to be engineered through mutations to improve their physicochemical properties (*8-10*). These modifications may aim to enhance stability under extreme temperature (*11*) or pH conditions, or to increase enzymatic activity and specificity. The process of optimizing proteins for such industrial applications typically involves iterative cycles of mutation, screening, and selection - a labor-intensive and time-consuming endeavor.

As computational simulations and related technologies continue to advance, various software tools have emerged to enhance protein thermostability, including Rosetta (*12*), ABACUS (*13*), and FoldX (*14*), which employ physical or statistical potential functions. While these computational methods often provide relatively accurate stability predictions, their capacity to predict protein biological activity is limited. Typically, modifying the biological activity of proteins requires long-term (~years) meticulous experimental research into their working mechanisms, which is the primary method of rational protein design. However, mechanistic research is time-consuming and labor-intensive, and it increasingly fails to meet the modification needs of many important industrial enzymes commonly used in everyday applications. In recent years, deep learning has been extensively applied in protein engineering. Large-scale protein language models (*15-19*), such as those utilizing self-supervised learning of protein sequences to understand protein sequence semantics and grammar, have demonstrated appreciable predictive performance for protein fitness (*20*), even in zero-shot settings (*19, 21, 22*). A zero-shot setting here means the model can predict the mutation sites of a protein to improve its properties without relying on any prior experimental mutagenesis data. However, the prediction of most protein language models pre-trained on extensive protein sequences often do not achieve sufficient high accuracy for protein stability, which is crucial for protein engineering (*23*). Other supervised deep learning methods exhibit high accuracy in predicting protein fitness but rely on high-throughput experiments to generate hundreds or even thousands of data points (*24, 25*). This approach is not practical for many proteins due to resource limitations. In this study, we used a comprehensive dataset comprising 96 million sequence-host bacterial strain optimal growth temperatures (OGT) (*26*). The optimal growth temperature of host bacterial strains has been shown to strongly correlate with information such as protein optimal enzymatic activity temperature and melting temperature (*27*). Leveraging this dataset, we developed a deep learning-based methodology, termed PRIME, which stands for Protein language model for Intelligent Masked pretraining and Environment (temperature) prediction. During its pretraining process, PRIME utilizes a masked language modeling (MLM) task, inspired by the transformer-based language models (*28*). This task involves artificially modifying protein sequences based on the natural probability distribution of amino acids, then attempting to restore the sequences to their original state. This procedure enables PRIME to learn and comprehend the semantic and grammatical features inherent in protein sequences. Alongside this, PRIME capitalizes on a multi-task learning paradigm to capture the temperature traits associated with these sequences. This approach fosters an inherent predisposition in PRIME to assign higher scores to protein sequences exhibiting enhanced temperature tolerance and conforming to natural biological principles. PRIME is trained with the objective of predicting optimal growth temperatures (OGTs) across a wide range of bacterial strains. As a result, PRIME naturally correlates higher scores with sequences more likely to contribute to robustness and survivability in varied environmental conditions, including extreme temperatures. Therefore, PRIME proves particularly proficient in the design and optimization of industrial enzymes and proteins that often require high-temperature tolerance and resilience for practical applications. Our model has demonstrated much better predictive performance compared to other state-of-the-art models in forecasting thermostability (change of $T_m$) and fitness prediction of mutated protein sequences.

To further evaluate the efficacy of our model, we applied it to five distinct proteins and subjected the results to wet-lab experimental validation. The proteins studied included LbCas12a, T7 RNA polymerase, creatinase, non-natural nucleic acid polymerase and the variable domain of the heavy chain of a nano-antibody against growth hormone (VHH). Without any prior experimental mutagenesis data, we used the PRIME model to select several top-ranking single-site mutants for experimental testing. Our results revealed that over 30% of these mutants displayed notable improvements in the physicochemical properties, such as thermostability, catalytic activity, antigen-antibody binding affinity, or even the nonnatural properties, e.g., the ability to polymerize non-natural nucleic acid or resilience to extreme alkaline conditions.

Protein engineering for various pharmaceutical and industrial applications is confronted by two major challenges. The first is the identification of beneficial single-site mutations, and the second is the combination of multiple single-site mutations into a deep mutant. The latter becomes particularly challenging as it is often observed that combining two positive single-site mutations often results in a two-site mutant with inferior performance compared to each single-site mutants before the combination. As shown in Ref (*29*) in high-throughput screening of green fluorescence protein, the probability of observing the negative epistatic effect, where the fluorescence intensity of a mutant combining two single-site mutations is worse than the linear addition of the fitness of the two before combination, is ~100 times higher than that of observing the positive epistatic effect. Building upon the foundational framework provided by PRIME, we introduce a multi-site stacking strategy based on the PRIME model. For example, in the case of T7 RNA polymerase, after three rounds of AI-experiment iterations with fewer than 100 mutants in total, we successfully developed a mutant with 12-site mutations that surpasses the thermostable counterpart offered by the leading commercial biotechnology company, New England Biolabs. We also conducted similar strategy on LbCas12a, which contains multiple domains and 1228 amino acids. After three rounds of AI-experiment iterations with fewer than 100 mutants, we achieved an 8-site mutant with the best thermostability to date, whose $T_m$ is 6.5 °C higher than the wild type while maintaining comparable or higher trans-cleavage activity at the desired condition.

Furthermore, in the case of T7 RNA polymerase and LbCas12a, we found PRIME can automatically combine the negative single-site mutations from different functional domains into a multi-site deep mutant to further improve the fitness of the latter. This could be a very important discovery as it opens a route for protein engineers as they now can make use of negative mutations to improve the fitness of proteins. These negative mutations, which are more common than positive ones, were traditionally pre-excluded in conventional protein engineering.

# Results

**PRIME Architecture**

PRIME is a pre-trained model based on the Transformer architecture(*30*), as illustrated in the Fig. 1A. PRIME consists of three main components. The first is the encoder module for sequence feature extraction, which is a Transformer encoder model to extract the latent representation of the sequence. The second component is the masked language modeling (MLM) module, which is designed to prompt the encoder to learn the contextual representation of amino acids. Meanwhile, the MLM module can also be applied in mutant scoring. The third component is the Optimal Growth Temperature (OGT) prediction module, which can predict the OGT of the organism in which the protein is located, based on the latent representation. The model and training details of PRIME are described in the **Methods**.

1) The pre-training objectives of PRIME

There consists of three learning objectives of PRIME: the Masked Language Modeling (MLM) objective, the Optimal Growth Temperature (OGT) prediction objective, and the correlation objective. The details of these objectives are as follows:

1.1) Masked language modeling

Masked Language Modeling (MLM) is often used as a pre-training method for sequential data representation. In this objective, noised protein sequences serve as the input, wherein parts of tokens are masked as '<mask>' or substituted with alternative tokens. The training objective is to reconstruct these noised tokens. This approach facilitates the model's ability to capture dependencies among amino acids as well as contextual information along the sequence. The details can be found in the **Methods**. Moreover, we can use this reconstruction process to score mutations.

1.2) Optimal growth temperature (OGT) prediction

The second training objective is optimized under supervised conditions. We utilize a dataset containing 96 million protein sequences annotated with OGT to train the PRIME model. The input of this objective is protein sequence, and the OGT module generates a temperature value ranging between 0 and 100°C. Notably, the OGT and MLM modules operate with a shared encoder. This architecture enables the model to simultaneously capture amino acid contextual information and temperature-related sequence characteristics (Fig.1B).

1.3) Correlation objective

We introduce a learning objective to align these two metrics to facilitate feedback from the predicted OGTs to the MLM scores. For a group of single-site mutant sequences, the OGT prediction module outputs their OGTs, and the MLM module scores these mutants. Subsequently, we maximize the Pearson correlation between these mutant scores and predicted OGT values, serving to align the mutant OGT with their corresponding mutant scores. The goal of this objective is the maximization of the Pearson correlation coefficient. We use Pearson correlation as our learning objective due to its differentiable properties (for backpropagation), in contrast to the non-differentiable of Spearman correlation.

We have conducted experiments using MSE loss to align the MLM and OGT predictions (Table S1). We found that this approach yielded inferior results compared to using Pearson correlation as a loss function. The possible reason is that MSE loss aligns the MLM and OGT values for a single sequence, resulting in unstable loss for individual data, and the absolute value of the MLM score holds limited significance for us. In contrast, correlation loss is calculated for a set of mutated sequences and better reflects the relative magnitude of values within a set, which aligns more closely with our specific application scenario of protein engineering and evaluating a set of mutated data.

2) Zero-Shot Single-Site Mutation Scoring.

Models trained with the Masked Language Modeling objective can output the likelihood of amino acids appearing at a specific position based on the surrounding context. We use this to score single-site mutations. Given a mutation, we treat the amino acid in the wild-type protein as a reference and compare its likelihood to that of the mutated amino acid. The mutations are then scored using the log-odds ratio at the mutated position. (See Fig. 1C; the details can be found in the **Methods**).

3) Augmentation of single-site mutation prediction performance in PRIME through fine-tuning on homologous sequences via the MLM learning objective.

While PRIME exhibits commendable performance in zero-shot mutant effect prediction, we observed that additional unsupervised fine-tuning of the language modeling module on homologous protein sequences of target proteins yields improved results, without adding supervision from experimental data. Explicitly, for the fine-tuning process, we deploy homologous sequences of the proteins of interest as an unsupervised dataset, optimizing both the encoder and masked language modeling (MLM) modules of PRIME and ESM2-650M. Evaluation results substantiate that this method improves PRIME's and ESM-2's predictive accuracy for mutant effect prediction.

**PRIME outperforms state-of-the-art methods in predicting fitness of mutated protein sequence**

We conducted a comparison of the zero-shot prediction capacity on thermostability between our model, PRIME, and several current state-of-the-art models, including deep learning models ESM-1v(*21*), ESM-2(*19*), MSA-transformer(*17*), Tranception-EVE(*31*), CARP(*32*), MIF-ST(*33*), SaProt(*34*), Stability Oracle(*35*) as well as the traditional computational method, GEMME(*36*) and Rosetta(*12*). Notably, among these methods, MIF-ST, SaProt and Rosetta incorporates protein structure information, whereas the others rely solely on protein sequence. Our analysis utilized a dataset derived from MPTherm(*37*), FireProtDB(*38*) and ProThermDB(*39*), featuring single-site mutations in proteins with $\Delta T_m$, i.e., changing of melting temperature as compared to the wild type, collected under the same experimental pH and ensuring a minimum of 10 data points per protein, which amassing a total of 66 assays. Concurrently, the analysis also incorporated assays from Deep Mutational Scanning (DMS), specifically those housed within ProteinGym(*31*). ProteinGym presents a meticulously constructed substitution benchmark, characterized by the experimental delineation and assessment of approximately 2.5 million missense variants. These variants are dispersed across 217 distinct DMS assays and encompass a range of protein properties including, but not limited to, enzymatic catalysis, binding affinity, stability, and fluorescence intensity. Such a comprehensive assembly of missense variants within the substitution benchmark of ProteinGym provides a robust and expansive repository, thereby facilitating the nuanced evaluative study of the myriad documented missense variants. This repository thus serves as a valuable asset for the systematic examination and interpretation of the diverse and intricate landscape of protein mutations and their associated properties.

These comprehensive datasets enabled a systematic investigation of the impact of specific mutations on protein fitness and thermostability, supporting the development and validation of advanced predictive models such as PRIME. The comparison provides valuable insights into the relative performance of different modeling approaches and highlights the potential of PRIME for predicting protein mutations in a zero-shot setting. The results are illustrated in Fig. 2A and Table S2. As can be seen, PRIME demonstrates better performance than all the other methods in predicting protein fitness and stability. In the ProteinGym benchmark, PRIME outperforms the second-best model, SaProt, registering a score of 0.486 against 0.457 (p=1e-4, Wilcoxon). In the $\Delta T_m$ dataset, PRIME's performance significantly surpasses the next model, Stability Oracle, with scores of 0.437 and 0.412 respectively (p=9e-3, Wilcoxon). We also compared PRIME with other methods in the dataset of Stability, which refers to ProteinGym-stability, a sub dataset of ProteinGym. PRIME still outperform all of the other methods. It's crucial to note that the OGT employed by PRIME isn't a direct representation of protein $T_m$. Instead, a correlation exists between them(*27*). There are some enzymes from thermophiles turn out to be not very thermostable(*40*). However, even when leveraging the slightly imprecise OGT as a stand-in for protein sequences' $T_m$ attribute, PRIME markedly outshines models that don't incorporate OGT. For instance, the similar-architecture counterpart, ESM-2, achieves only 0.330 in the $\Delta T_m$ dataset. We posit that PRIME's performance would witness a significant boost with access to a vast dataset of accurate $T_m$ values for natural proteins. These findings underscore PRIME's potential in protein engineering endeavors, particularly in crafting protein sequences with enhanced thermostability and other fitness attributes. Across the board, PRIME outclasses both traditional computational strategies and other deep learning models, underscoring its unparalleled effectiveness.

Recent efforts, such as SaProt, which integrates protein structural information into protein language model (PLM) models, show enhanced prediction capabilities on stability. However, SaProt and other structural models, including MIF-ST or Stability Oracle, require protein structure data as input, which inherently carries noise and is limited by the availability of high-precision structures either from wet-lab experimental resolution or predictions like those from AlphaFold. This makes their application somewhat restricted. PRIME, which only requires sequence input, has already outperformed the current leading model SaProt when using the latest complete version of ProteinGym (217 datasets) as a benchmark. Importantly, as a purely sequence-based model, PRIME not only substantially improves prediction capabilities within stability datasets compared to other

PLMs, such as ESM-2, but also achieves superior performance in non-stability datasets, particularly those involving activity, as shown in Table S2.

In addition to the zero-shot assignment, we also tested the representational capacity and transferability of PRIME. Specifically, we conduct supervised fine-tuning on two temperatures related downstream tasks with global finetuning (Fig.1B). As the pretraining of PRIME incorporates the optimum growth temperature of the bacterial where the protein lives in, it is anticipated that PRIME can also perform better in predicting other properties of proteins associated with temperature. As exhibited from Fig. 2B to 2E (Table S3), PRIME also outperforms other supervised methods in the task of predicting the melting temperature ($T_m$) of a native protein and its optimal enzymatic activity temperature ($T_{opt}$). Considering the importance of $T_m$ and $T_{opt}$ in protein, PRIME's ability to rapidly label a large volume of protein native sequences with thermal properties, using only sequence input, is of notable utility for native protein annotation engineering in practical applications.

Furthermore, we delved deeper into understanding the individual contributions of the three core modules within PRIME: the OGT prediction module, the mask language modeling module (MLM), and the correlation term. Our findings, detailed in Table S1, highlight that relying solely on either the OGT prediction or the MLM module leads to a dip in PRIME's performance. Among these, the MLM module stands out as having the most pronounced effect across all zero-shot benchmarks. The OGT module plays a pivotal role in $\Delta T_m$ prediction, with the standard PRIME achieving a score of 0.437, in contrast to PRIME/-OGT which scores 0.362 (p=3e-2, Wilcoxon). Similarly, the correlation term significantly influences $\Delta T_m$ prediction, with PRIME/-correlation registering a score of 0.429 (p=4e-2, Wilcoxon). In the context of the ProteinGym benchmark, both the OGT and correlation terms continue to exert a significant influence. This finding highlights the significance of combining the OGT prediction, MLM, and correlation modules in the PRIME model to achieve optimal performance. The synergistic effect of these three modules allows the model to better understand the complex relationships between protein sequences and their thermostability properties, ultimately resulting in improved predictive capabilities. The integration of all these modules in the PRIME model ensures a more comprehensive understanding of the protein sequence information, which in turn contributes to its superior performance compared to other state-of-the-art models.

Further, we assessed PRIME's performance in other supervised protein engineering tasks. Specifically, in the FLIP benchmark(*41*), which consists of 12 tasks, PRIME leads in all of these tasks over ESM-1b, ESM-1v, ESM-2 and CARP, demonstrating its strong extrapolation capability, particularly in predicting high-complexity mutational effects (Table S4). We note that, among the 12 tasks in FLIP, two of them (AAV and GB1) correspond to predicting the fitness of the multi-site deep mutants when knowing the fitness of the constituent single-site mutations, which is crucial for identifying the final product in the protein engineering. One plausible explanation for this capability is that during PRIME’s pre-training, there is an alignment between the token-level MLM and the sequence-level OGT attributes of mutant sequences. This alignment allows the model to learn the thermal properties of native sequences and the thermal stability ranking of mutant sequences. Since protein thermal stability, binding affinity, and other extremophilic tolerances follow similar physical principles reflecting structural stability, PRIME exhibits superior extrapolation capability in tasks related to native protein thermal stability (Meltome) and mutated protein binding affinity (AAV, GB1) within the FLIP benchmark. This is why PRIME demonstrates a stronger performance in these tasks compared to the ESM series. Moreover, in the Meltome (*42*) dataset task of FLIP, which involves predicting the $T_m$ of human-derived proteins, PRIME, integrated with OGT information, consistently surpassed models with similar architectures like ESM-2. This indicates that even though PRIME's pre-training process only learned the OGT information of bacterial-derived protein sequences, it still excels in predicting the $T_m$ temperature attributes of proteins from other species. This demonstrates PRIME's generalizable capabilities.

**Wet-lab experimental testing of PRIME-designed single-site mutants of various proteins for different engineering purposes**

In practical applications of protein engineering, the prevailing approach involves identifying positive single-site mutations that enhance the protein's performance (making it more active or more stable), and then combining them to form multi-site mutants with desired properties probably through a greedy search method (*25*). Thus, the successful identification of these positive single-site mutations forms the cornerstone of successful protein engineering. To further substantiate the effectiveness and generosity of our methodology, we tested the PRIME model on designing single-site mutant for five distinct proteins, namely LbCas12a, T7 RNA polymerase, creatinase, non-natural nucleic acid polymerase (Tgo-D4K) and the variable domain of the heavy chain of a nano-antibody against growth hormone (VHH). Briefly, we fine-tuned PRIME on a set of 30,000 homologous sequences for each target protein, sourced from the Uniclust30 database (*43*). This fine-tuning was executed with five distinct random seeds for each target protein. By averaging the prediction outcomes from these five models for single-site saturation mutations, we generated a single-site mutation score table for every protein. PRIME was then utilized to rank all single-site mutants within the landscape, based on the likelihood of the mutated sequences relative to their wild-type counterparts (refer to the Mutated protein sequence scoring strategy). Subsequently, we selected top 30-45 mutants from outside the 6-angstrom range of the catalytic active sites or binding pockets for further experimentation. Considering that mutations within the catalytic active site or binding pockets could profoundly affect the protein’s function, direct mutation of the active site presents both risks and opportunities (*44*). In this study, we adopted a conservative approach aimed at averting potential drastic disruption to the protein’s catalytic capabilities. Notably, we were initially unsure about the specific effects of PRIME's suggested mutations on the properties of the five proteins under study. However, each protein had distinct enhancement needs, either in stability or activity. For the five distinct proteins, the engineering objectives varied: for LbCas12a, T7 RNA polymerase and creatinase, the goal was to enhance thermostability; for non-natural nucleic acid polymerase, the target was to accelerate the polymerization rate of non-traditional nucleic acids, specifically 2'-fluoroarabino nucleic acid; and for VHH, the objective was to increase stability in highly alkaline pH conditions (pH>13). Comprehensive outcomes of these experiments are elaborated upon in the subsequent sections.

PRIME can be used to rank mutants based on both activity and stability for single mutants. However, from the ablation study of PRIME, we found that the zero-shot performance with only the OGT module (PRIME/-MLM) is quite poor in both the ProteinGym benchmark and $\Delta T_m$. Therefore, we do not use the OGT module to select single-site mutations for stability. Instead, we suggest using the LLM likelihood of PRIME, obtained when predicting OGT as an additional pretraining task. Drawing on previous research experience of biologists (*7, 44*), we can choose mutations located on the surface of the protein to improve protein stability, and mutate amino acids around the protein pocket to enhance protein catalytic activity.

**LbCas12a**

It is well known that engineering proteins with multiple functions is challenging due to the trade-offs between different protein properties (*45*). Moreover, these multi-domain proteins often have substantial conformational differences between their functional states and their crystal structures, which poses a substantial challenge to traditional rational design methods that rely on structure. Thus, we sought to challenge our model with a large, multi-domain protein whose activity requires cross-talk between multiple functional domains. We engineer the $T_m$ of *Lachnospiraceae bacterium* Cas12a (LbCas12a). Cas12a is an RNA-guided endonuclease belonging to the type V-A CRISPR-Cas system (*46*). LbCas12a contains 1,228 amino acids with multiple functional domains (Fig. 3A). During the catalytic process, crRNA guides Cas12a to bind to and cleave double-stranded DNA substrates. Upon target DNA recognition, the REC lobe of Cas12a undergoes conformational changes to unleash its trans-activity to cleave non-specific single-stranded DNA (*47*). This feature makes LbCas12a particularly useful in in-vitro diagnostic applications (*48*). We used PRIME to perform a round of single-site mutation prediction and tested 30 single-site mutations, of which 9/30 single-site mutants had a $T_m$ not lower than the wild type (V936F, I976L, S962K, M957L, M456I, L59K, Y549K, G49K, C1090D) (Fig. 3B).

**T7 RNA polymerase**

T7 RNA polymerase is a monomeric enzyme derived from T7 bacteriophage, comprising a total of 883 amino acids. Since its initial utilization in RNA synthesis in the early 1980s, T7 RNA polymerase has become a crucial tool in the fields of molecular biology and genetic engineering (*49*). It is now commonly employed in various applications such as in vitro transcription experiments, mRNA vaccine production (*50*) and isothermal amplification detection techniques (*51, 52*), etc. However, T7 RNA polymerase also presents some application drawbacks. For example, it produces immunostimulatory by-products, such as double-stranded RNA (dsRNA), during the transcription process (*53*), which necessitates complex purification processes for mRNA vaccine production. Recent studies have indicated that increasing the reaction temperature to above 48°C effectively reduces the by-products (*54*). Nevertheless, the wild-type T7 RNA polymerase unfolds at temperatures around 45°C, resulting in decreased enzymatic activity and an inability to transcribe the desired target products at higher temperatures. Therefore, there is a critical need to enhance the thermal stability of T7 RNA polymerase.

In this study, we employed PRIME to predict mutation sites in T7 RNA polymerase and directly selected the top 45 single-site mutants for subsequent experimental verification. As depicted in Fig. 3D, our experimental results indicate that 57.7% (26 out of 45) of the mutants have a $T_m$ value higher than the wild type (Y846R, C125A, H772K, S606V, V687E, F481W, C216L, N601E, A881F, P657K, S633P, K642G, M306K, A703T, P476E, A468F, T375K, I217L, R792M, S397W, P865L, C515P, W797L, S430P, L446F, Q786L).

**Creatinase**

Creatinase, a dimeric proteinase, is widely employed in enzymatic assays for measuring creatinine levels (*55, 56*). It is primarily derived from microorganisms such as *Pseudomonas*., *Bacillus*., and *Alcaligenes.* Creatinase is crucial in medical diagnostics and plays a role in quantifying creatinine in serum and urine (*57*). Elevated creatinine levels indicate impaired kidney or muscle function. Nevertheless, the optimal catalytic temperature for creatinase typically falls within the range of 30-40 °C, which constraints both the industrial and clinical diagnostic applications. Enhancing the thermal stability of creatinase not only improves the efficiency of clinical creatinine detection but also facilitates enzyme production, storage, and transportation.

Here, we employed the PRIME model to predict single-site mutations in creatinase obtained from *Alcaligenes faecalis* (*58*). At the end 28 single-site mutants were selected for experimental validation. As depicted in Fig. 3F, 32% (9 out of 28) of the mutants exhibited improved thermal stability (Q151V, H193Y, V283L, A180K, Y310L, E170T, S19L, H74Q, D17V).

**Non-natural nucleic acid polymerase**

Tgo is a DNA polymerase that has been identified in the thermophilic bacterium *Thermococcus gorgonarius*, which was isolated from a geothermal vent in New Zealand (*59*). Tgo has been found to accurately replicate 2'-fluoroarabino nucleic acid (FANA), a genetic polymer with 2'-fluoroarabino residues in deoxyribonucleotides (*60, 61*). However, Tgo DNA polymerase can only catalyze the synthesis of FANA on the DNA template at a rate of ~15 nt/min (*61*), which is much lower than that of Tgo for DNA synthesis (~400 nt/min) (*62*), limiting the application of FANA as a substitute for DNA in information storage (*63*), disease treatment (*64, 65*) and other fields. The evolution of an XNA polymerase necessitates a comprehensive evaluation of not only binding affinity but also catalytic activity and processivity. This is due to the unique chemical and biophysical properties of XNA, which differ from those of DNA and RNA, making prediction by traditional in silico methods challenging. Furthermore, the distinct sugar pucker of XNAs may result in conformational structures of XNA that differ from those of DNA, RNA, and nucleic acids modified in bases, thereby influencing the polymerase's recognition of XNA. Consequently, the in-silico prediction and direct evolution of XNA polymerase remain formidable challenges. To date, the evolution of XNA polymerases has relied solely on random mutation methods in vitro, such as the compartmentalized self-tagging (CST) method. Vitor et al. constructed a high-throughput mutation library and conducted at least two rounds of screening to identify the currently fastest FANA polymerase, Tgo-D4K (TgoT: L403P, P657T, E658Q, K659H, Y663H, E664K, D669A, K671N, T676I) (*66*). The polymerase was able to extend FANA on the DNA template at a rate of ~80±27 nt/min, while the rate of DNA extension on the DNA template was reduced to 16±3 nt/min (*62*). However, the synthesis rate of Tgo-D4K for FANA is still lower than that of Tgo for DNA synthesis. Therefore, methods are required to modify existing polymerases to screen for polymerases with higher FANA synthesis rates.

In the present study, we commenced our investigation with Tgo-D4K as the starting point. Utilizing PRIME, we systematically screened potential mutation sites across various domains of Tgo-D4K. Ultimately, we selected 27 promising mutations for subsequent experimental validation. The polymerase kinetic profiling (PKPro) strategy was used to detect the FANA synthesis rate of the mutants as previously described (*62*). The experimental results (Fig. 3H) showed that more than 40% (12 out of 27) of the mutants had a higher FANA synthesis rate (P716G, R460E, I528A, H659E, K465E, A546V, I471E, D29V, Y481G, T55L, A217P, I693W), and the single-site (I693W) mutation was identified, which can notably increase the extension rate to ~3.2-fold of that of the Tgo-D4K enzyme.

**VHH**

VHH antibody is the antigen binding fragment of heavy chain only antibodies (*67*). Due to the advantages of small size, monomer state, robust structure and easy tailoring, VHH has been employed as an important tool in medical research and clinical antibody drug development (*68*), which have been developed as an affinity ligand to selectively purify biopharmaceutical, for example prothrombin, tetrabromobisphenol A, intercellular adhesion molecule 1 and so on (*69-71*). In the practical production of biological products, the most widely used method of clean-in-place (CIP) is 0.5 M NaOH cleaning for 24 hours. Hence, VHH antibodies used for biopharmaceutical purification need mutational engineering to tolerate the harsh alkaline condition, which is rarely seen in nature (*72, 73*).

In this study, we employed our PRIME model to predict mutation sites for an VHH antibody against growth hormone which we select from an immunized camelid. The top 29 mutants were chosen for further testing, 11 of 29 (~38%) mutants enhanced stability after incubation at 0.3 M NaOH for 24 h, as shown in Fig.3J (A57D, P29T, A15P, V113D, P117Q, R20T, R110E, T58K, D114Y, W112F, L12K). Among these, the A57D mutation displayed a remarkable twelve-fold enhancement in alkali tolerance. Besides, ~31% (9 out of 29) of the mutants show increased affinity for antigen before the alkaline treatment (P29T, A15P, A57D, P117Q, Q83D, R20T, T119V, L12K, V113D).

**Benchmark of different strategies for selecting single-site mutations**

To evaluate the efficiency of PRIME and the strategy of our single-site mutation selection. We incorporated a benchmark comparison for different strategies of selecting single-site mutations. We conducted comparisons both in silico and through wet lab experiments. From the ProteinGym dataset, we used a subset with saturated single-site mutation data (comprising five datasets with wild-type sequence identity to the PRIME pre-training dataset below 30%) for this analysis. We compared the top-15 single-site mutations selected by four different strategies, which include: 1) the strategy method in this paper, using homologous sequences of the target protein to finetune the PRIME model; 2) finetuning ESM-2 on the same homologous sequences; 3) the ESM vote strategy from the Ref (*74*); 4) random single mutations. Our single-site selection strategy consistently outperformed the other methods across three evaluation metrics: the number of positive single-site mutations, the maximum fitness, and the median fitness of the mutants. The specific results are presented in Table S5. Furthermore, we compared the performance of the top-15 single-site mutations selected by different methods through wet lab experiments. We limited our comparison to two proteins: T7 RNA polymerase and a non-natural nucleic acid polymerase Tgo-D4K. We validated the top-15 single-site mutations selected by our strategy, the ESM vote strategy, the strategy of fine-tuning ESM-2 on homologous sequences, and the strategy of scoring saturated single-site mutations with Rosetta for unfolding free energy. Rosetta scores protein saturated single-point mutations by ranking based on predicted values of the unfolding free energy. The energy function used to calculate this unfolding free energy includes all energy terms referenced in the literature (*75*). The results, shown in Fig. 4A and 4B (detailed in Table S6), demonstrate that our strategy method's selected single-site mutations comprehensively outperform those selected by other strategy methods.

**Enhanced Multi-site Mutagenesis through PRIME-Driven Protein Engineering**

Traditional protein engineering and directed evolution techniques often employ an incremental approach, reminiscent of greedy algorithms, accumulating mutations from single-site mutants to construct multi-site variants. Such a strategy, while prevalent, is prone to pitfalls, notably converging to local optima. Specifically, the most effective multi-site mutant doesn't always emerge from the aggregation of the top-performing single-site mutants. Harnessing the capabilities of PRIME, we unveil an advanced multi-site mutation stacking strategy. This purely data-driven method evaluates the entire landscape of $2^N$ potential mutants (where $N$ represents the count of single-site mutations available for combination), bypassing the pitfalls of traditional directed evolution that might settle for local optima through incremental mutations. Our strategy simplifies the prediction of high-performing multi-site mutations, reducing the need for extensive experimental iterations, as depicted in Fig. 1D. Importantly, our methodology includes a zero-shot prediction pipeline based on homologous sequences with PRIME fine-tuned for specific proteins. Previous studies have indicated that protein language models fine-tuned on homologous sequences can achieve substantial better performance in low-$N$ scenarios (*76*). A comparative analysis with the ftMLDE method from Reference (*77*), using simulated directed evolution on the GB1 dataset, demonstrates that our PRIME-based workflow more effectively identifies multi-site mutants with enhanced max fitness (Fig. 4C) or mean fitness (Fig. 4D). We examined both the top-K sampling used by PRIME and the tiered sampling employed by ftMLDE (*77*). Our findings indicate that the iterative multi-point mutation strategy based on PRIME outperforms ftMLDE in terms of both maximum and average fitness across multiple rounds of iteration. Where the results of top-K sampling were comparable to tiered sampling, with top-K sampling showing a slight advantage (Detailed results were shown in Table S7). This in-silico directed evolution was conducted 100 times, with each iteration comprising two rounds. In each round, the top-50 mutants or tiered-50 samples identified from the preceding round were used as the training dataset for the following round, employing an MLP layer as the regression model for ESM-2 and PRIME to score the whole rest mutants. For the implementation of ftMLDE, we executed the code as described in Ref (*77*), and utilized MSA-transformer as the variant encoding model, and the regression module is ensemble of ARDRegression, BaggingRegressor, and KNeighborsRegressor.

**LbCas12a**

In the case of LbCas12a, we trained the PRIME model on all the 30 single-site mutation data points and predicted the $T_m$ of multi-site mutation combinations. The top 10 scored mutants were then selected from each of the two to four-site mutation pools for the second-round experimental validation. In the third-round stability evolution, the 30 multi-site mutants were added to the training set to further finetune PRIME. We then selected the top 5 mutants from each of the three-, four-, five- and six-site mutant collection and top 10 mutants in total from the seven-, to ten-site mutant collections were selected for experimental characterization.

As shown in Fig. 5C, 17 out of 30 multi-site mutants in the second-round exhibited higher $T_m$ than the wild type. Furthermore, all the 30 multi-site mutants in the third-round had a higher $T_m$ than the wild type. The best mutant was an 8-site mutant (R2-26) with $T_m$ of 48.15°C, which is 6.25 °C higher than wild type (details can be found in supplemental materials).

Interestingly, in the second round of $T_m$-enhancing positive multi-site mutations, many of the multi-site mutations recommended by PRIME contain negative single-site mutations ($T_m$ decrease). For example, C10L in the R1-3 (C10L; S962K) mutation has a $T_m$ lower than the wild type, but it participates in the formation of this double-site mutation with a $T_m$ higher than both two single-site mutations. Moreover, the three-site mutation R1-15 (C10L; S962K; I976L) formed by adding the C10L mutation based on the R1-9 (S962K; I976L) double-site mutation also has a $T_m$ higher than the previous double-site mutation. Furthermore, the positive multi-site mutations containing C10L consist of mutations from different functional domains. As shown in Fig. 5A, for instance, C10L is in the WED-I domain of cas12a, while S962 and I976 are in the RuvC-II domain. This demonstrates the remarkable generalization ability of PRIME, which has learned the epistatic effects between different mutations from different domains with only the information of sequence, and can combine negative single-site mutations into excellent multi-site mutations. This is challenging to achieve with traditional directed evolution methods, which employ an incremental approach, reminiscent of greedy algorithms, accumulating mutations from single-site positive mutants to construct multi-site variants. It is unlikely to directly combine negative single-site mutations into multi-site mutations.

**T7 RNA Polymerase**

Taking T7 RNA polymerase as another example, we built upon the foundation of previously identified single-site mutations, and employed the PRIME model, fine-tuned with homologous sequences of T7 RNA polymerase, to perform supervised regression prediction tasks (details can be found in **Methods**).

We utilized the $T_m$ data from all single-site mutations in the first round, including the wild-type protein, as the training set and then employed the trained PRIME models to predict the multi-site mutation sequences. Subsequently, from sequences with 2-4 mutation sites, we selected five sequences each and ten sequences for 8 mutation sites that had the highest predicted $T_m$, resulting in a total of 25 multi-site mutants for the second round of wet-lab validation. As shown in Fig. 5D, after two rounds of mutagenesis, all 25 multi-site mutants exhibited a $T_m$ higher than the wild-type. The standout mutant had 8 mutation sites, R1-21 (Q786L; S430P; W797L; P657K; N601E; L446F; P476E; T375K), with its $T_m$ being 7.4 °C higher than the wild-type.

However, compared to the commercial thermostable T7 RNA polymerase (Hi-T7, $T_m$ = 56.8 °C) available from New England Biolabs (NEB), our 8-site mutant still has a $T_m$ that is 4°C lower. To acquire a mutant with higher $T_m$, we further tested 10 additional single-site mutants from the first round of zero-shot prediction by PRIME, as shown in Fig. 5D. We then combined the data from all single-site mutants and previous multi-site mutants, a total of 80 mutants, to train the PRIME model. Subsequently, we used the trained PRIME model to directly predict the $T_m$ of multi-site mutants range from 9 to 14-site mutations formed by these single-site mutations. And the top-15 multi-site mutants predicted by PRIME were selected into the following wet-lab testing. Remarkably, 5 out of 15 deep mutants showed unambiguous higher $T_m$ as compared to Hi-T7, with the best mutant of 12 mutation sites (Q786L; S430P; L446F; S606V; K642G; S633P; I217L; S397W; L534V; A124N; G618E; L665D), whose $T_m$ was 12.8 °C higher than that of wild type. Notably, the enzymatic activity of the five most thermostable mutants were also higher than the wild type, as illustrated in Fig. 5D.

Furthermore, we found this 12-site mutant contains several negative single-site mutations, such as A124N, G618E and L665D. When applied to the wild type, this mutations would lead to decrease in $T_m$, as shown in Fig. 5D. The amalgamation of negative mutations poses a formidable challenge, as these mutation sites are often preemptively excluded from further combinations to form deep mutants in conventional protein engineering due to the paucity of domain knowledge on their effective utilization. Given that negative mutations are far more common than positive ones, our finding that protein LLMs can make use of them as ingredient to form better deep mutants could be exciting to the protein engineering community for further mechanism and industrial applications.

Without any prior experimental data or high throughput screening technique, after three rounds of mutagenesis and wet-lab validation of 95 mutants, we successfully obtained a T7 RNA polymerase variant with up to 12-site mutations that surpasses the commercial enzyme. This achievement not only attests to the precision and efficiency of PRIME's single-site prediction and multi-site stacking but also highlights its potential in notably reducing the financial overheads associated with wet experiments. This accomplishment remains elusive in the realm of traditional protein engineering and rational design.

# Discussion

We present PRIME, an advanced deep learning approach that masterfully leverages an extensive dataset encompassing sequence-host bacterial strain optimal growth temperatures. By tailoring a masked language model (MLM) for Optimal Growth Temperature (OGT) prediction, PRIME astutely captures the semantic, grammatical, and temperature-related nuances of protein sequences. Rigorous in silico evaluations consistently underscore PRIME's preeminence over other leading models, including ESM-1v, ESM-2, MSA-transformer, Tranception-EVE, CARP, MIF-ST, SaProt, Stability Oracle, GEMME and Rosetta, in predicting thermostability and the overall fitness of protein mutants. Through PRIME, we have crafted five proteins with single-site mutations, achieving substantial enhancements in their physicochemical attributes, with a commendable success rate of over 30% among the 30-45 AI-conceptualized mutants. This highlights PRIME's transformative potential in the realm of protein engineering.

Historically, protein engineering strategies have pivoted around either directed evolution or rational design. The former, while effective, hinges on high-throughput experimental screenings, making it resource-intensive in terms of both time and capital. For numerous pivotal proteins, the practicality of high-throughput experimental methodologies is debatable, rendering low-throughput assays a more viable alternative. Conversely, rational design demands an in-depth comprehension of the biophysical attributes pertinent to the target protein's operational mechanism. With a profound understanding of this mechanism, rational design can occasionally identify high-performing mutants with limited wet experiments. Yet, for many proteins with limited mechanistic insights or for modifications of unconventional activities, such as the polymerization activity towards non-natural nucleic acids highlighted in our study, rational design often encounters limitations. In these scenarios, AI-centric predictions, epitomized by PRIME, stand out. Without necessitating extensive wet experimental data or a deep understanding of the protein's modus operandi, PRIME offers invaluable insights, streamlining the protein engineering trajectory.

Traditional protein engineering often adopts a strategy akin to greedy algorithms, incrementally accumulating mutations from single-site to multi-site mutants. While effective, this process can be labor-intensive and time-consuming. Moreover, it occasionally results in suboptimal outcomes, as the optimal multi-site mutant does not necessarily comprise the most beneficial single-site mutants. Our model, PRIME, introduces a paradigm shift in this field. It offers a refined strategy for multi-site mutation accumulation, overcoming the limitations of conventional tactics and expediting the creation of superior multi-site protein mutants. Remarkably, PRIME can automatically group negative single-site mutations into a deep mutant, notably enhancing its fitness. This discovery could be pivotal, opening a pathway for protein engineers. They can now utilize negative mutations, which are more prevalent than positive ones and were previously excluded in traditional design, to enhance the fitness of proteins. By reducing the reliance on exhaustive experimental screenings, computational tools like PRIME could revolutionize the protein engineering landscape, potentially expanding the range of proteins amenable to skilled engineering. This approach holds promise for a wide array of applications in pharmaceutical and industrial sectors.

Furthermore, PRIME's versatile modeling framework holds promise for diverse predictive tasks, such as deducing the melting temperature ($T_m$) or the optimal enzymatic activity temperature ($T_{opt}$) of indigenous proteins. PRIME streamlines the prerequisites for protein modifications, facilitating enhancements in protein stability and activity, eliminating the need for comprehensive mechanistic probes. Additionally, PRIME's multi-task learning modality, which aligns OGT with MLM, considerably boosts the model's predictive accuracy on temperature-associated downstream tasks when juxtaposed with other training techniques. Moreover, this does not compromise its predictive efficacy on tasks unrelated to temperature. This suggests that while enhancing the model's predictive capability for specific tasks, this pretraining method also maintains the model's generalization capability on other unrelated tasks. This pretraining approach could pave the way for a fresh learning paradigm, embedding specialized domain insights into foundational AI frameworks, and could be instrumental in bridging the gap between deep learning and conventional scientific wisdom. Remarkably, PRIME's predictive prowess extends to pinpointing mutation sites that bolster protein properties, even those seldom observed in nature. Instances include fortifying antibody resilience in extreme alkaline environments or amplifying a polymerase's polymerization velocity on non-native nucleic acids, underscoring PRIME's universal applicability in protein engineering.

# Methods

## Details of PRIME Architecture

PRIME consists of a common Transformer-based encoder and two different components: one for performing masked language modeling (MLM) pretraining and another for pretraining optimal growth temperature (OGT) prediction. In this section, we first introduce the common Transformer encoder, followed by a detailed description of the MLM module and the OGT prediction module.

**1) Transformer Encoder**

For the Transformer encoder, we use the same architecture of the ESM-2, a widely used Transformer-based pre-trained language model. Compared to the standard Transformer model architecture, it replaces the absolute position embedding with rotary position embedding and utilizes the trick of pre-layer normalization like Roberta, and the activation unit is a GELU function rather than ReLU. We also utilize Flash attention to accelerate the training and inference. The code can be found in our code repository. Conceptually, the Transformer encoder acts as a parameterized transformation function, converting a protein sequence into a sequence of dense vectors:

$$(h_1, h_2, \ldots, h_L) = Transformer(x_1, x_2, \ldots x_L)$$

Here $L$ is the length of the protein sequence, $(x_1, x_2, \ldots x_L)$ represents the discrete one-hot encoded amino acids of the protein sequence, and the continuous vectors $(h_1, h_2, \ldots, h_L)$ are the outputs of the Transformer encoder, representing the protein sequence in latent space.

**2) Masked Language Modeling Module.**

This module is also the same as ESM-2 architecture. This module acts as a reverse function of the Transformer encoder, mapping a sequence of hidden vectors into the one-hot encodings of protein sequences. During pre-training, the MLM module is learned to recover the noised protein sequence. The noised sequence is generated heuristically from the original sequence by randomly masking 20% of the tokens in a protein sequence. Of these masked tokens, 70% are replaced with a special <mask> token, accounting for 14% of the entire sequence. Additionally, 20% of the masked tokens, or 4% of the entire sequence, are substituted with amino acids. These substitutions are based on their natural occurrence frequencies in the UniProtKB database, ensuring that more common amino acids have a higher likelihood of being chosen. The objective of this task is to let the encoder to understand the relationships between words and to learn the contextual information necessary for understanding the primary structure of protein sequences. More importantly, it has been shown that the probability distribution generated by the model for a given masked position in a protein sequence, over all possible amino acids, has a positive correlation with the mutant score(*78*). The mutant score is a measure of how likely it is that a given amino acid substitution at that position will result in a functional change in the protein. The fact that the probability distribution generated by the model is correlated with the mutant score indicates that the model has learned to capture important features of protein sequences, such as the effects of amino acid substitutions on protein function. Formally, the masked language modeling module is a point-wise parameterized function, converting a sequence of dense vectors into a sequence of probability distribution on the protein sequence.

$$p_j = MLM\left(h_j\right) j \in noised\ postions,$$

where $j$ denotes the noised position and $h_j$ is the latent representation while $p_j \in R^{20}\ is$ probability distribution (20 is the vocab size).

**3) OGT Prediction Module.**

The original masked language model for natural languages is actually join-trained with an additional supervised task that learns to decide if two given sentences follow each other or not. However, this supervised part is usually ignored in protein-based models. To fill this gap, we added a supervised module to our model to learn how to predict the optimal growth temperature (OGT) of the organism to which a protein belongs. This module contains an attention-based pooling layer, two multi-layer perceptron (MLP) layers and a residue connection. The attention pooling layer takes the latent representations of the protein sequence $(\boldsymbol{h}_1, \boldsymbol{h}_2, \ldots, \boldsymbol{h}_N)$ as input, subsequently utilizes an projection-softmax layer to compute the weights and produces a weighted vector $\boldsymbol{c}$:

$Attention(\boldsymbol{h}_1, \boldsymbol{h}_2, \ldots, \boldsymbol{h}_N)$:

$$(\boldsymbol{h}_1, \boldsymbol{h}_2, \ldots, \boldsymbol{h}_N) \leftarrow LayerNorm(\boldsymbol{h}_1, \boldsymbol{h}_2, \ldots, \boldsymbol{h}_N)$$

$$s_i = \frac{\exp(\boldsymbol{W}\boldsymbol{h}_i + \boldsymbol{b})}{\sum_{n=1}^{N} \exp(\boldsymbol{W}\boldsymbol{h}_n + \boldsymbol{b})}\ i = 1,2, \ldots, N$$

$$\boldsymbol{c} = \sum_{n=1}^{N} s_i \boldsymbol{h}_i$$

, where $\boldsymbol{W}$ and $\boldsymbol{b}$ are the learnable parameters of the attention pooling layer.

Then an MLP layer with two fully connected layers and GELU activation is employed to transform the weighted vector $\boldsymbol{c}$. The first fully connected layer maps $\boldsymbol{c}$ to the same dimension as the Feed-Forward Network (FFN) layer of the Transformer, which in our implementation is four times the size of the hidden layer. The second fully connected layer maps the output of the first layer back to the original dimension. Between the first and second fully connected layers, there is a GELU activation function. Additionally, there is a residual connection between the output of the second fully connected layer and the output of the attention layer:

$$\boldsymbol{r} = FC_2\left(\boldsymbol{g}\left(FC_1(\boldsymbol{c})\right)\right) + \boldsymbol{c}$$

, where $FC_2$ and $FC_1$ are learnable fully-connected layers, $g$ is the GELU activation function. Especially, the output vector $\boldsymbol{r}$ can be viewed as a representation feature of the whole sequence, which can be used in the transfer learning for downstream tasks.

Finally, another MLP layer with two fully-connected layers and a Tanh activation function are employed to learn to map the sequence representation $\boldsymbol{r}$ to the OGT of the protein sequence:

$$y = FC_4(tanh(FC_3(\boldsymbol{r}))) \; y \; is \; OGT \in \boldsymbol{R}$$

where $FC_3$ and $FC_4$ are trainable fully-connected layers. We utilize the mean square error (MSE) criterion as the loss function.

**Zero-shot Prediction of the Effects of Single-point Mutations**

According to (*18, 21*), Protein language models, which are trained using the masked language modeling objective, are capable of predicting the likelihood of an amino acid occurring at a specific position in a protein based on the surrounding context. This prediction ability can be utilized to evaluate sequence mutant effects. Fig. 1C shows how to predict the mutant effect using the Masked language modeling module. For a given mutation, the amino acid in the wild-type protein serves as a reference state. The effect of the mutation is ascertained by comparing the predicted probability of the mutated amino acid against that of the original (wild-type) amino acid. Formally, the effect of the mutation is quantified through the log odds ratio at the mutated position, as:

$$Score(i, m|w) = logP(x_i = m|X) - logP(x_i = w|\boldsymbol{X})$$

where $Score(i, m|w)$ represents the score of the single-point mutant, where the $i_{th}$ wild-type amino acid $w$ has been mutated to mutant type $m$. And $\boldsymbol{X} = (\boldsymbol{x}_1, \ldots, \boldsymbol{x}_L)$ denotes the entire wild-type sequence, where $\boldsymbol{x}_i$ indicating the amino acid at position $i$ and $L$ is the sequence length. Note that this process can also be applied to multi-point mutant effects, where the fitness value of mult-site mutations can be considered as the sum of the fitness of its single-site mutations. This method is used to evaluate multi-point mutants in the ProteinGym benchmark.

# Training Details

**Pretraining**

As shown in Fig. 1A, PRIME incorporates three distinct loss functions as optimization objectives during pre-training: MLM Loss, OGT prediction loss, and the correlation loss. Below, we provide detailed formula of these three functions. The training and validation curves during the pretraining process are depicted in Fig. S1.

**Masked Language Modeling Loss**: To compute the MLM loss, we employ the Cross-Entropy loss. For each masked amino acid in a protein sequence, the model computes the probability distribution over its vocabulary (20 naturally occurring amino acids) and compares it to the actual amino acid (AA) distribution (represented as a one-hot vector). The loss is the negative log-likelihood of the

correct amino acid:

$$L_{MLM} = -\sum_{i} \log P(AA_i^{true})$$

Where $P(AA_i^{true})$ represents the predicted probability of the true amino acid.

**OGT Prediction Loss:** This loss function is used to quantify the difference between the predicted OGT and the actual OGT. We employ the Mean Square Error (MSE) as the loss function, which can be expressed as follows:

$$L_{OGT} = \frac{1}{N}\sum_{i}\left(T_i^{pred} - T_i^{true}\right)^2$$

Here, $N$ represents the number of training samples, $T_i^{pred}$ signifies the predicted OGT, and $T_i^{true}$ represents the true OGT.

**Correlation Loss:** This loss function aligns the mutation scores generated with the predicted OGT of mutations. Given a protein sequence $S$, we randomly generate $N$ single-point mutants $\boldsymbol{M} = (M_1, M_2, \dots, M_N)$. Utilizing the Masked Language Model (MLM) Module, we can obtain MLM scores $\boldsymbol{S} = (S_1, S_2, \dots, S_N)$ of these $N$ mutants. And the OGT Module is used to predict the temperatures $\boldsymbol{T} = (T_1, T_2, \dots, T_N)$ for these mutants. Pearson correlation coefficient is then utilized to align $\boldsymbol{T}$ and $\boldsymbol{S}$, with the specific formula given by:

$$L_{Corr} = 1 - \frac{cov(\boldsymbol{S}, \boldsymbol{T})}{\sigma_S \sigma_T}$$

, where $cov(\boldsymbol{S}, \boldsymbol{T})$ represents the covariance between $\boldsymbol{S}$ and $\boldsymbol{T}$, and $\sigma_S$ and $\sigma_T$ are the standard deviations of $\boldsymbol{S}$ and $\boldsymbol{T}$ respectively.

The final loss is the sum of three model losses. We observed that the OGT Prediction Loss has a significantly different magnitude compared to the other two losses, with values ranging from 0-1000 initially and stabilizing at 0-100 later. To maintain numerical stability, we multiplied this loss by 0.01.

**Implement Details**

We utilized PyTorch to implement PRIME. The transformer encoder is comprised of 33 layers and 20 attention heads, with 650 million parameters and an embedding size of 1280. The learning rate was set to $1 \times 10^{-4}$. The micro-batch size per GPU is 4096 tokens, and the gradient accumulation steps are 32. The models were trained for 200k update steps on 8×A100 80G GPUs. After pre-training, the root mean square root of the OGT prediction task was 3.5 on the 50,000 held-out validation set, and the perplexity of masked language modeling reached 3.52. The average error of the correlation loss during pretraining reached 0.1623. We initialize all layers of the transformer encoder and MLM module from (*19*).

**Alternating Training**

Due to the disparate input requirements of the three loss functions—MLM Loss operates on noised protein sequences, OGT Loss on complete sequences, and Correlation Loss on $N$ random single-point mutants of the sequence—we use alternating training strategy to optimize these distinct objectives. Specifically, we utilize Mini-batch Gradient Descent with the Adam optimizer to train the model, alternating tasks with each mini-batch iteration. The training regimen is delineated in the Python and PyTorch-style pseudocode in Table S8. After training, we compared the predicted and actual OGT, as shown in Fig. S2.

**Effect of different weight of the multitask loss function to the performance of zero-shot prediction**

We have explored the design of the multitask loss function with varying weights to address the relative amount of data or task difficulty. To minimize computational costs, we randomly selected 500,000 sequences from the full pre-training dataset of 96 million entries to serve as our training dataset for these ablation studies. Each combination of loss weights was selected using a grid search from the list [0.01, 0.05, 0.5, 1, 2], resulting in a total of 125 combinations. We discovered that a 1:1:1 weight ratio presents an optimal setting for the zero-shot mutation prediction task on the ProteinGym and $\Delta T_m$ datasets. The specific results are documented in the Table S9.

**Fine-tuning MLM Module on Homologous Sequence**

To improve the performance of PRIME and ESM-2 in zero-shot mutant effect prediction, we explore enhancing it through fine-tuning on homologous sequences with only training on the masked language modeling. Fine-tuning on homologous sequences involves adapting a pre-trained model to a specific protein by leveraging the knowledge gained from similar protein sequences (*78*). Our approach applies this fine-tuning strategy to the ProteinGym or $T_m$ dataset. Using Jackhammer, a renowned sequence comparison tool, we identified homologous sequences of proteins within these datasets from the Uniclust30 database (*43*). For proteins with over 30,000 homologous sequences, the first 30,000 sequences were selected. Conversely, for those with fewer than 30,000, all sequences were retained for fine-tuning. The fine-tuning process employed the same hyperparameter settings as in the pre-training phase of MLM module. Specifically, the noised sequence is generated heuristically from the original sequence by randomly masking 20% of the tokens in a protein sequence. Of these masked tokens, 70% are replaced with a special <mask> token, accounting for 14% of the entire sequence. Furthermore, 20% of the masked tokens, corresponding to 4% of the entire sequence, are substituted with amino acids based on their natural occurrence frequencies in the UniProtKB database, ensuring that more common amino acids are more likely to be chosen. Our objective in fine-tuning on these sequences is to harness the shared attributes among homologous proteins, thereby enhancing mutation effect predictions. This tailored approach aims to optimize the pre-trained model for specific protein contexts, offering a promising avenue for enhanced predictive accuracy.

**Transfer Learning of PRIME on Temperature Related Benchmark and FLIP.**

PRIME is trained on both temperature-related supervised and unsupervised tasks. To assess the transfer representational ability of PRIME, we utilize a $T_m$ prediction benchmark and another optimal catalytic temperature prediction ($T_{opt}$) benchmark. The assessments were executed utilizing the encoder component and the OGT Module; All of the parameters of the transformer encoder can be fine-tuned. The batch size is set to 256 and the learning rate is set to 0.0001 in the Adam optimizer. Moreover, the model was subjected to early stopping, with a patience setting of 20 epochs and the max number of training epochs is set to 200 epochs. The loss function is also MSE. To ensure robustness, the experiments were executed in five-fold cross-validation. There is no information in the test set that was used during training and validation. The mean of the results was used as the final performance metric, and the variance was utilized for the error bars.

**Transfer Learning of PRIME on Supervised Mutant Effect Prediction.**

PRIME can also be applied in supervised mutant effect prediction tasks, which is used in our strategy for generating multi-site mutants (Fig. 1D). Given a training set of mutated sequences with experimental fitness labels. We can utilize PRIME to learn on the training set and further predict the fitness of new mutated sequences. In this task, we only use the transformer encoder and OGT module, while the MLM module is dropped. Except for the parameters of the last two fully connected layers, $FC_3$ and $FC_4$, in the OGT module, which are re-randomized, the rest remain frozen, which is called Regression Module in Fig. 1D. We also use MSE as a loss function to learn how to minimize the predicted mutated fitness and the true fitness. During this training, the learning rate is 1e-4, and the batch size is 16. The training epochs are dynamically decided. We begin by splitting the dataset into five folds. In each iteration, we use four folds for training and the remaining one for validation. We track the number of epochs needed for the validation set across these iterations, resulting in five epoch counts. The final number of training epochs for the entire dataset is determined by averaging these five epoch numbers. In the final training phase, we do not use a validation set; instead, we train on the entire dataset using the previously determined number of epochs. After training, the model can be utilized to predict the fitness of unseen mutated sequences. For proteins without any labeled mutant data, we first employ the zero-shot capability of PRIME, referred to as PRIME (Zero-shot), to select the top-*K* single-point mutants. These mutants are then experimentally labeled. In the first round of design, we use this labeled data to train PRIME, called PRIME (Round 1). This trained model is used to predict fitness scores for multi-site mutants combined from all single-site mutants identified in the zero-shot round, and the top-*K* mutants are selected based on these scores. We then experimentally determine the fitness of these top mutants. The labeled multi-site mutant data from this first round is added to the initial training set, and PRIME is retrained on this updated set, called PRIME (Round 2). Using PRIME (Round 2), we predict fitness scores for all multi-site mutants combined from both the zero-shot and Round 1, selecting the top-*K* mutants. We then experimentally determine the fitness of these top mutants. If the results do not meet the requirements, we further add these labeled top mutants to the training set and repeat the process.

## Dataset

### Pretraining Dataset

By integrating publicly accessible data from Uniprot and protein sequences from metagenomic projects (*79-81*), we have curated ProteomeAtlas, a vast repository of natural protein sequences containing 4.7 billion entries. We filtered these sequences, retaining only those that are full-length. Further, we employed MMseqs2 to process these sequences, setting a sequence identity threshold of 50% for redundancy reduction. This enabled us to identify and annotate sequences corresponding to optimal growth temperatures (OGT) (*26*) for bacterial strains. Ultimately, we annotated 96 million sequences in this manner, providing a rich resource for exploring protein sequence-temperature relationships.

### Benchmark Datasets for Zero-Shot Mutation Scoring

The dataset utilized for changes in melting temperature ($\Delta T_m$) was sourced from MPTherm (*37*), FireProtDB (*38*) and ProThermDB (*39*), ensuring that all experiments were conducted under the same pH conditions. The ProteinGym dataset was cited from Reference (*31*). Datasets and data split for predicting melting temperature ($T_m$) and optimal enzymatic activity temperature ($T_{opt}$) of native protein sequences were drawn from Reference (*82*).

**Different Strategies of selecting single-site mutations for different engineering purposes**

PRIME can be used to rank mutants based on both activity and stability for single mutants. However, from the ablation study of PRIME, we found that the zero-shot performance with only the OGT module (PRIME/-MLM) is quite poor in both the ProteinGym benchmark and $\Delta T_m$. Therefore, we do not use the OGT module to select single-site mutations for stability. Instead, we suggest using the LLM likelihood of PRIME, obtained when predicting OGT as an additional pretraining task. Drawn on the past research experience of biologists (*7, 44*), one can choose mutations located on the surface of the protein to improve protein stability while not alerting much the activity and mutate amino acids around the protein pocket to enhance protein catalytic activity. This empirical knowledge can be utilized in a specific protein engineering assignment, which might further increase the success rate.

## Engineering of high stability or activity in five proteins

### Prediction of single-site mutations by PRIME

First, we employed Jackhmmer to identify sequences homologous to each target protein within the Uniclust30 database(*43*). For proteins with a bounty of over 30,000 homologous sequences, we randomly cherry-picked a subset of 30,000 for the fine-tuning of the PRIME model. On the other hand, for proteins boasting fewer than 30,000 homologous sequences, we incorporated all available sequences into the fine-tuning process. This fine-tuning was executed across five iterations, each initiated with a distinct random seed, for every target protein. By amalgamating the predictive outcomes from these five distinct model parameters for single-site saturation mutations, we synthesized a comprehensive mutation scorecard for each protein. Mutants that showcased scores surpassing that of the wild type in the scorecard were earmarked as potential candidates. In the final phase, we meticulously handpicked approximately 30-45 mutants, ensuring they were situated beyond the 6-angstrom radius of pivotal regions like the catalytic active sites or binding pockets, to pave the way for subsequent experimental evaluations.

### T7 RNA polymerase

**Preparation of T7 RNA polymerase variants**

The T7 RNA polymerase (Uniprot ID: P00573) gene and its mutants' gene were cloned into the pQE-80l expression vector and transformed into *E. coli* BL21(DE3) cells. The cells were cultured in LB media until reaching an OD600 of approximately 0.6-0.8, followed by induction with 1 mM IPTG for a 6-hour growth period at 37°C. After collection, the bacteria were resuspended in a binding buffer (50 mM Tris-HCl pH 8.0, 300 mM NaCl, 3 mM imidazole, 0.1 mM EDTA) and lysed via sonication. The resulting lysate underwent centrifugation at 4°C and 12,000 rpm for 30 minutes. The lysate was then applied to a Ni-NTA gravity column and washed with a washing buffer (50 mM Tris-HCl pH 8.0, 300 mM NaCl, 10 mM imidazole, 0.1 mM EDTA, 10% glycerol). Elution was performed using an elution buffer (50 mM Tris-HCl pH 8.0, 300 mM NaCl, 250 mM imidazole, 0.1 mM EDTA, 10% glycerol). Concentration was achieved using a final ultrafiltration buffer (50 mM

Tris-HCl pH 8.0, 100 mM NaCl, 0.1 mM EDTA), and the T7 RNA polymerase was diluted with a storage buffer (50 mM Tris-HCl pH 8.0, 100 mM NaCl, 0.1 mM EDTA, 1 mM DTT, 75% glycerol) (*83*).

**Thermal melt measurements**

The protein staining agent, SYPRO Orange, was added to a final concentration of 5×, and the protein sample (~0.2 mg/ml) was mixed in an eight-row PCR tube. Each sample was prepared in a final volume of 20 µl and tested in triplicate. Denaturation curves were generated using a PCR instrument (Analytik Jena qTower3) equipped with appropriate optical filters (FAM470nm and ROX625nm for excitation and emission, respectively). The temperature was incrementally increased by 0.5 °C steps from 25 to 65 °C, with a 5-second hold for equilibration at each temperature step. The thermal unfolding curves were analyzed by fitting the Boltzmann equation to approximate the $T_m$ (*58*).

***In vitro* transcription assays**

The *in vitro* transcription reaction buffer was prepared, which contained 200 mM HEPES (pH 7.5), 30 mM $MgCl_2$, 20 mM DTT, 0.4 U/µL RNase inhibitor, 5 mM NTPs mix, and 100 nM iSpinach DNA template (*84*). The buffer was incubated at 52 °C for 10min and then 0.04 mg/ml T7 RNAP was added to initiate the reaction. After the mixture incubated for 1 hour, 100 mM EDTA was added to stop the reaction. Finally, 100 µM DFHBI was introduced, and fluorescence was measured with excitation at a wavelength of 470 nm and emission at 512 nm.

## Creatinase

**Preparation of Creatinase variants**

The creatinase (Uniprot ID: Q9RHU9) gene was cloned into a pET-28a expression vector and transformed into *E. coli* BL21(DE3) cells. The cells expressing creatinase were cultivated in LB medium supplemented at a temperature of 37°C while agitating the culture at 220 rpm. To induce the expression of creatinase, when the OD600 value of the culture reached 0.8-1.0, isopropyl-β-D-thiogalactopyranoside (IPTG) was added at a final concentration of 1 mM. The cells were then further cultured at a reduced temperature of 18°C for a duration of 16 hours. After collecting the cells, they were resuspended in a binding buffer (25 mM Tris-HCl pH 8.0, 200 mM NaCl, 20 mM imidazole) and subjected to sonication for cell disruption. The resulting lysate was centrifuged at 4°C and 12,000 rpm for 30 minutes, and the supernatant was collected. The supernatant was loaded onto a pre-equilibrated Ni-NTA gravity column, and protein elution was performed using an imidazole gradient ranging from 20 to 200 mM. The purity of the fractions obtained was analyzed using SDS-PAGE.

The fractions containing the purified target protein were combined and desalted using an ultrafiltration unit. The purified protein was then concentrated and stored in 1× PBS at a temperature of -80°C to maintain its stability and activity. (*58*)

**Differential scanning fluorimetry (DSF)**

The thermal stability testing was also carried out using a PCR instrument (Analytik Jena qTower3). All proteins were diluted in 1× PBS to a final concentration of 0.3 mg/ml and mixed with SYPRO Orange at a final concentration of 5× in an eight-row PCR tube. The protein unfolding process was initiated by subjecting the samples to a thermal treatment ranging from 25 to 85°C (with a temperature increment of 0.5°C/step) with each step holding for 5 seconds.

Subsequently, the thermal unfolding curves were obtained, and the data were analyzed using the Boltzmann equation to determine the $T_m$. (*58*)

**Activity measurements**

Creatine could be hydrolyzed by creatinase into urea and creatinine. The resulting urea reacts with p-dimethylaminobenzaldehyde to form a yellow-colored dye. The concentration of urea can be determined by measuring the absorbance of the yellow dye at 435 nm using a spectrophotometer. (*58*) Consequently, the specific activity of the protein can be calculated. Here are the details of the experimental procedure:

1. Incubate a PBS buffer solution (280 μL) containing 100 mM creatine at 37°C for 5 minutes.
2. Incubate the mixture with 20 μL of protein solution (1 mg/mL) for 22 minutes.
3. Stop the reaction by adding p-dimethylaminobenzaldehyde solution (600 μL) prepared by dissolving 2 g of p-dimethylaminobenzaldehyde in 100 mL of dimethyl sulfoxide (DMSO) and 15 mL of concentrated hydrochloric acid.
4. Measure the absorbance at 435 nm using a spectrophotometer.

## VHH

**Protein expression and purification.**

The gene of the VHH was cloned into the pET29a plasmid with an N terminal His-tag. The expression plasmid was transformed into *E.coli* BL21(DE3) cells. A single colony of each recombinant *E. coli* strain was inoculated into 30 mL LB medium with 50 μg/mL kanamycin for seed culture at 37 °C for 12-16 h. The seed culture was transferred 10 ml to 1L LB medium with 50 μg/ml kanamycin at 37 °C 220rpm until the $OD_{600}$ value get 0.6-0.8. The culture was cooled to 16 °C and then induced with 0.5 mM IPTG for 20-24 h at 16 °C. Cells were harvested from the fermentation culture by centrifugation for 30 min at 4,000 rpm, and the cell pellets were collected for later purification. The cell pellets were resuspended in buffer A (20 mM $Na_2HPO_4$ and $NaH_2PO_4$, 0.5 M NaCl, pH 8.0) and then and lysed via ultra sonification. The lysates were centrifuged for 30 min at 12,000 rpm at 4 °C, after which the supernatants were subjected to Ni-NTA affinity purification with elution buffer (20 mM $Na_2HPO_4$ and $NaH_2PO_4$, 0.5 M NaCl, 250 mM imidazole, pH 8.0). The purity of the fractions obtained was analyzed using SDS-PAGE. The fractions containing the purified target protein were combined and desalted using an ultrafiltration unit. The purified protein was then concentrated and stored in buffer A with 10% glycerol at a temperature of -80°C.

**Protein dealed with alkaline**

Will-type and mutants of VHH were incubated at 0.3 M or 0.5 M NaOH for 3 h, 6 h and 24 h. Subsequently, hydrochloric acid was added to terminate the alkali treatment, the samples were stored at a temperature of -80°C.

**Alkaline pH stability test (ELSA)**

Ninety-six-well plates were coated with growth hormone protein at a desity of 5ng/well at 4 °C overnight. The plates were washed with 1 × PBS'T three times. Following blocking with 1% BSA in 1 × PBS'T at 25°C for 2 h. After washing three times with 1 × PBS'T, the plate were incubated with serial dilutions of VHH proteins 100μl/well (1:2,1:4.1:8,1:16,1:32,1:64,1:128,1:256,1:512,1:1024,1:2048) for 1h at 25°C. After washing three

times with 0.5% PBS'T, 100 μl/well HRP(1:5000) were added and incubated at 25°C for 1h. The plate were washed with 1 × PBS'T four times, a total of 100 μl/well TMB was added and incubated at 25 °C for 15 min in the dark. Finally, 100 μl/well 2 M $H_2SO_4$ was added to stop the reaction and absorbance was measured at 450 nm (TECAN, Swiss.).
The log(agonist) vs. response -- Variable slope (four parameters) curves were analyzed to calculate EC50 which determines the stability of VHH after alkaline treatment.

## Non-natural nucleic acid polymerase

### Polymerase expression and purification

Polymerases were expressed and purified as previously reported(*85*). Briefly, Tgo-D4K and its mutants' gene were cloned into the pGDR11 vector and transformed into *E. coli* BL21 cells. The cultures were grown in 50 mL LB medium containing ampicillin (100 μg/mL) at 37°C with shaking at 240 rpm until the $OD_{600}$ reached 0.6-1.0. Then, the cultures were induced by adding IPTG (0.5 mM) and incubated at 16°C with shaking at 240 rpm for 20 hours. The cells were harvested by centrifugation, and the pellet was lysed by sonication in buffer (10 mM Tris-HCl pH 8.0, 500 mM KCl, 10% glycerol). The lysate was centrifuged for 30 min at 13,300 rpm at 4 °C, and the clarified supernatant was heat for 1 h at 80 °C, then immediately cooling for 30 min on ice. The lysate was clarified again by centrifugation for 30 min at 4 °C and 13,300 rpm. 0.5% (v/v) polyethyleneimine was added to precipitate the nucleic acids, then the lysate was centrifuged for 30 min at 13,300 rpm at 4 °C. 60% (w/v) ammonium sulfate was added to precipitate the polymerase, after incubating for 1 h at 4 °C, it was centrifuged for 30 min at 13,300 rpm at 4 °C. Protein pellets were suspended in 4 °C pre-cooled buffer (10 mM Tris-HCl pH 8.0, 50 mM KCl, 10% glycerol). The supernatant was loaded onto Ni-NTA resin. All protein eluted at 200 mM imidazole was dialyzed in 4 °C buffer (10 mM Tris-HCl pH 8.0, 50 mM KCl, 10% glycerol). The purity of the fractions obtained were verified by SDS–PAGE and stored at -80 °C.

### Measurement of synthesis rates of polymerase

To measure the synthesis rates of the polymerase, kinetic measurements were performed as previously reported(*62*). Each measurement (10 μL) contained 1 μM of 30-mer template, 100 μM of each nucleotide triphosphate, 1× ThermoPol buffer, 2× LC Green Plus fluorescent dye, and 20 nM polymerase. Reactions were denatured for 3 min at 95 °C and extended for 30 min at 55 °C, with fluorescence intensity recorded at 6 s intervals. Fluorescence data for each polymerase were normalized and converted to nucleotides per polymerase. The synthesis rate was determined by linear fitting of nucleotides per polymerase over reaction time. The reported values are the average of three independent replicates.

## Lbcas12a

### Plasmids construction

LbCas12a mutants were constructed by overlap polymerase chain reaction (PCR) using a previous described pET28a plasmid harboring wild-type LbCas12a as the template and oligonucleotides carrying desired mutations. The expression plasmid contained a C-terminal 10 × His tag for downstream affinity purification. The recombinant plasmids were transformed into Escherichia coli Trelief 5α cells (Tsingke, China, Beijing). The sequences of all the plasmid constructs were confirmed via Sanger sequencing (Tsingke).

### Protein expression and purification

All the LbCas12a proteins were expressed in *Escherichia coli* BL21(DE3) cells cultured in Luria-Bertani (LB) medium supplemented with 50 μg/mL kanamycin. Single colonies were picked from the LB agar plates and grown in a starter culture overnight. The next day, the culture was inoculated into fresh LB medium supplemented with 50 μg/mL kanamycin at a ratio of 1:100 and incubated at 37 °C until OD600 reached 0.6. Protein expression was induced with 1 mM isopropyl β-D-1-thiogalactopyranoside (IPTG) at 37 °C for 4 h. The cells were harvested by centrifugation at 5,000 rcf for 15 min at 4 °C.

Collected cells were resuspended in lysis Buffer (pH 8.0) containing 100 mM sodium phosphate, 600 mM NaCl, 0.05% Tween-20, 30 mM imidazole, 1 mM dithiolthreitol (DTT) and 0.5 mM phenylmethylsulfonyl fluoride (PMSF). After disruption by sonication and centrifugation for 1 h at 12,000 rcf at 4 °C, HisPur Ni-NTA Magnetic Beads (ThermoFisher Scientific, Waltham, MA, USA) were used to purify proteins according to the manufacture's protocol. The harvested protein was concentrated into storage buffer containing 50 mM Tris-HCl, pH 7.5, 500 mM NaCl, 10% (v/v) glycerol and 2 mM DTT by Pierce Protein Concentrators (ThermoFisher) and stored at -80 °C.

**crRNA preparation**

All the DNA oligos used in this study were purchased from Tsingke Biotechnology Co. For crRNA preparation, in vitro transcription (IVT) template was generated by annealing a T7 promotor-carrying oligonucleotide with a complementary oligonucleotide containing antisense T7 promotor, crRNA direct repeat motif and spacer sequence. crRNA transcription was performed in a 30 μL reaction using the above IVT templates and HiScribe T7 Quick High Yield RNA Synthesis Kit (New England Biolabs) at 37 °C overnight. The residual DNA templates in the IVT reactions were removed by treatment with 0.08 U/μL deoxyribonuclease I (DNase I), and the RNA product was purified by TRIzol (Invitrogen).

**Differential scanning fluorimetry assays**

All the LbCas12a proteins were diluted to a final concentration of 0.5 mg/mL in reaction buffer containing 50 mM Tris-HCl, pH 7.5 and 500 mM NaCl and added into Standard Capillaries (NanoTemper). All the experiments were carried out at temperatures ranging from 20 to 95 °C with a heating rate of 1 °C per minute by using Prometheus NT.48 instrument and PR.ThermControl software (NanoTemper, Munich, Germany).

***In vitro* cleavage assays**

The Cas12a *trans*-cleavage reaction was performed as previously described (*86*) with minor modifications. Target DNA was PCR amplified from a plasmid via specific primers or generated by annealed oligonucleotides and then purified. Briefly, the reaction was carried out with 50 nM LbCas12a protein, 2.5 ng of substrate DNA, 100 nM crRNA, 0.5 mM DTT, 1.25 μM single strand DNA (5'-FAM-CCCCC-BHQ1-3'), and 1 × Buffer 2.1 (New England Biolabs) in a 10 μL reaction. Each sample was performed with three biological replicates and loaded on to 384-well plates. After incubation for 15 min at 42 °C, the fluorescence intensity was monitored using SpectraMax iD3 Multi-Mode Microplate Reader with an excitation wavelength of 485 nm and an emission wavelength of 535 nm. The fluorescence signal was recorded in 2 min interval and processed in subsequent analyses.

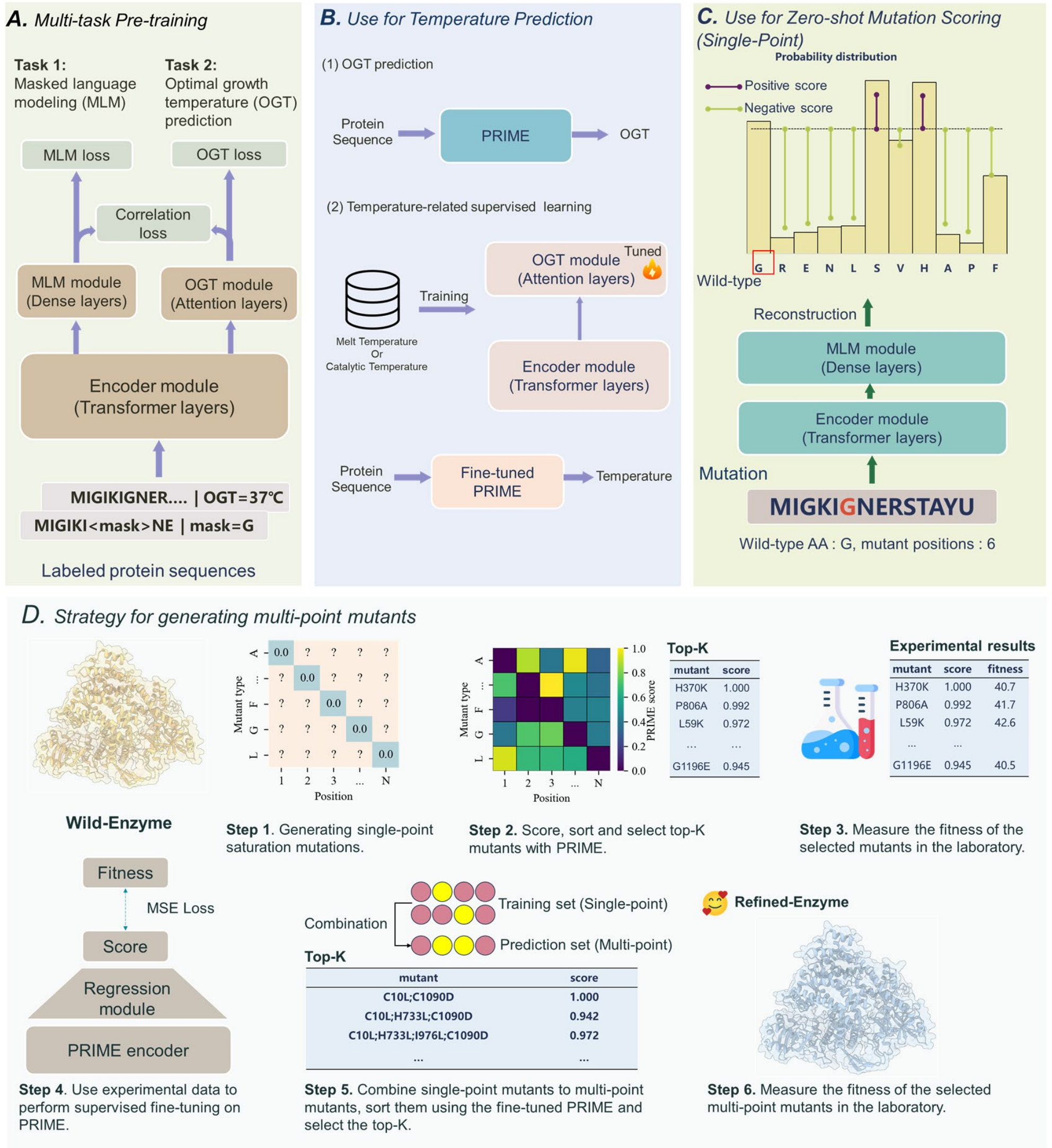

**Fig. 1. Overview of the PRIME architecture and its applications. A:** The architectural design of PRIME**.** PRIME incorporates a BERT-oriented transformer encoder, augmented by two domain-specific modules: one for Masked Language Modeling (MLM) and another for Optimal Growth Temperature (OGT) prediction tasks. The learning objectives comprise three distinct loss functions: MLM loss, quantified via cross-entropy; OGT loss, assessed through the mean squared error criterion; and correlation loss, evaluated by the inverse Pearson correlation coefficient. **B:** The use of PRIME for temperature prediction. PRIME can predict the optimal growth temperature of a protein sequence, and can be further fine-tuned with other temperature datasets (e.g., melting or optimal catalysis temperature). **C:** The use of PRIME for single-site mutation scoring. The wild-type sequence is reconstructed via the MLM module, generating a probabilistic distribution for amino acid identity at the mutation locus. The mutational impact is then quantified by the log-odds ratio between the mutated and wild-type amino acids. **D:** The strategy for generating multi-site mutants involves several steps. First, PRIME is used to assess the impact of single-site mutations, from which the top-K mutants are selected for experimental fitness evaluation in the laboratory. Subsequently, this experimental data serves as the training set to fine-tune PRIME. The fine-tuned model is then employed to predict the fitness of multi-site mutants. Based on these predictions, the top-K multi-site mutants are selected for further experimentation.

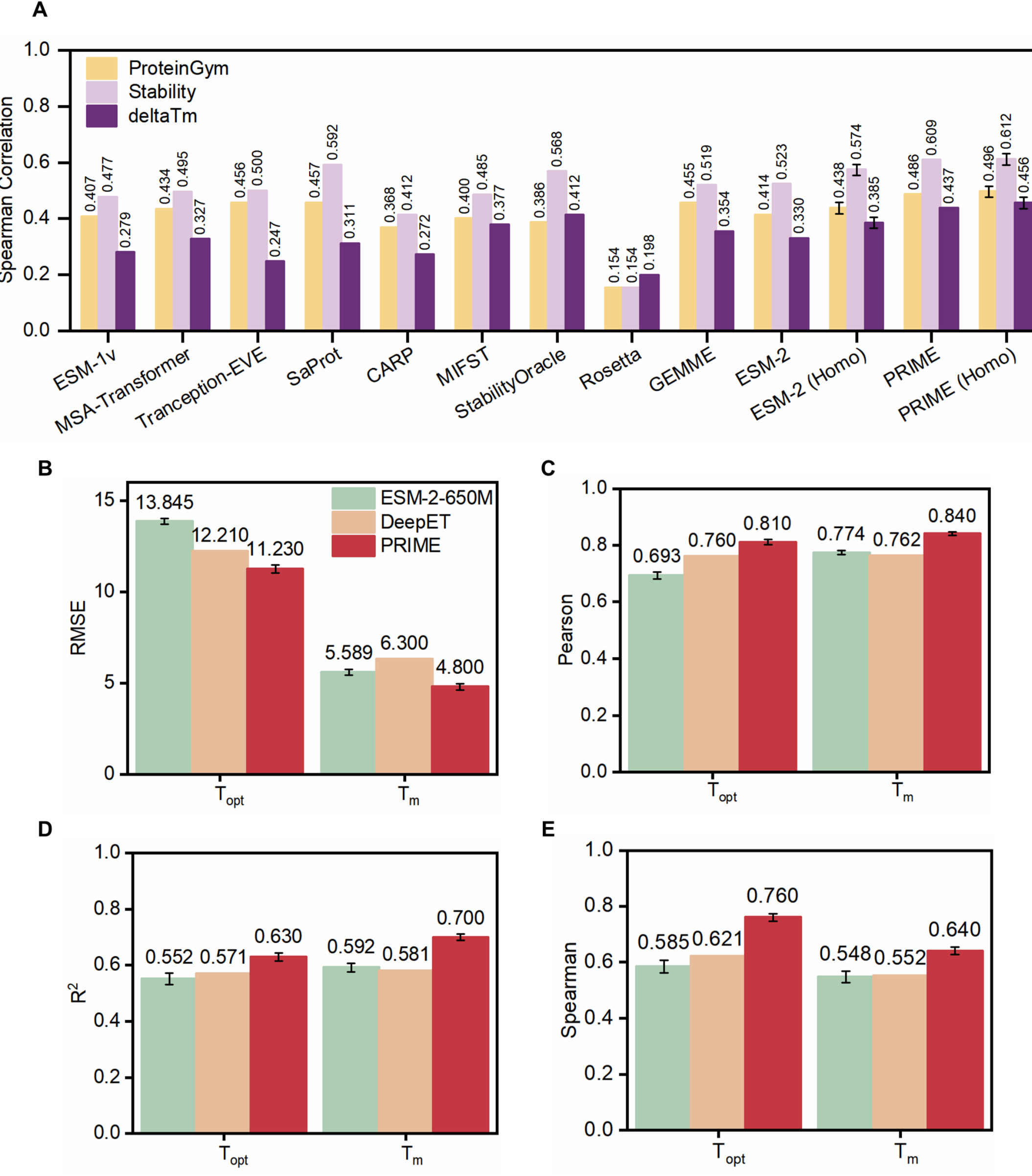

**Fig. 2**. **Comparison of the performance between PRIME and other methods.** A: Unsupervised model benchmarking on the $\Delta T_m$ and ProteinGym datasets. PRIME (homologous sequences) denotes the fine-tuning of the PRIME model using the MLM loss on homologous sequences of the target proteins present in either ProteinGym or $\Delta T_m$ datasets. B-E: Supervised prediction of $T_m$ (melting Temperature) and $T_{opt}$ (optimal enzymatic activity temperature). For the supervised benchmarks, we trained PRIME and ESM-2 with three different random seeds, while the results for DeepET were obtained from Ref(*82*). Four metrics are employed to gauge the models' accuracy and predictive ability: RMSE (B), Pearson correlation (C), $R^2$ (coefficient of determination, D) and Spearman correlation (E). The datasets and data split for $T_m$ and $T_{opt}$ are referenced from Ref(*82*). We obtained the wild-type protein structure from the Protein Data Bank and employed Alphafold2(*87*) to construct structures absent in PDB for the input to Rosetta and MIF-ST. The data points and the p-value tests associated with Fig. 2 are shown in Table S2 and S3.

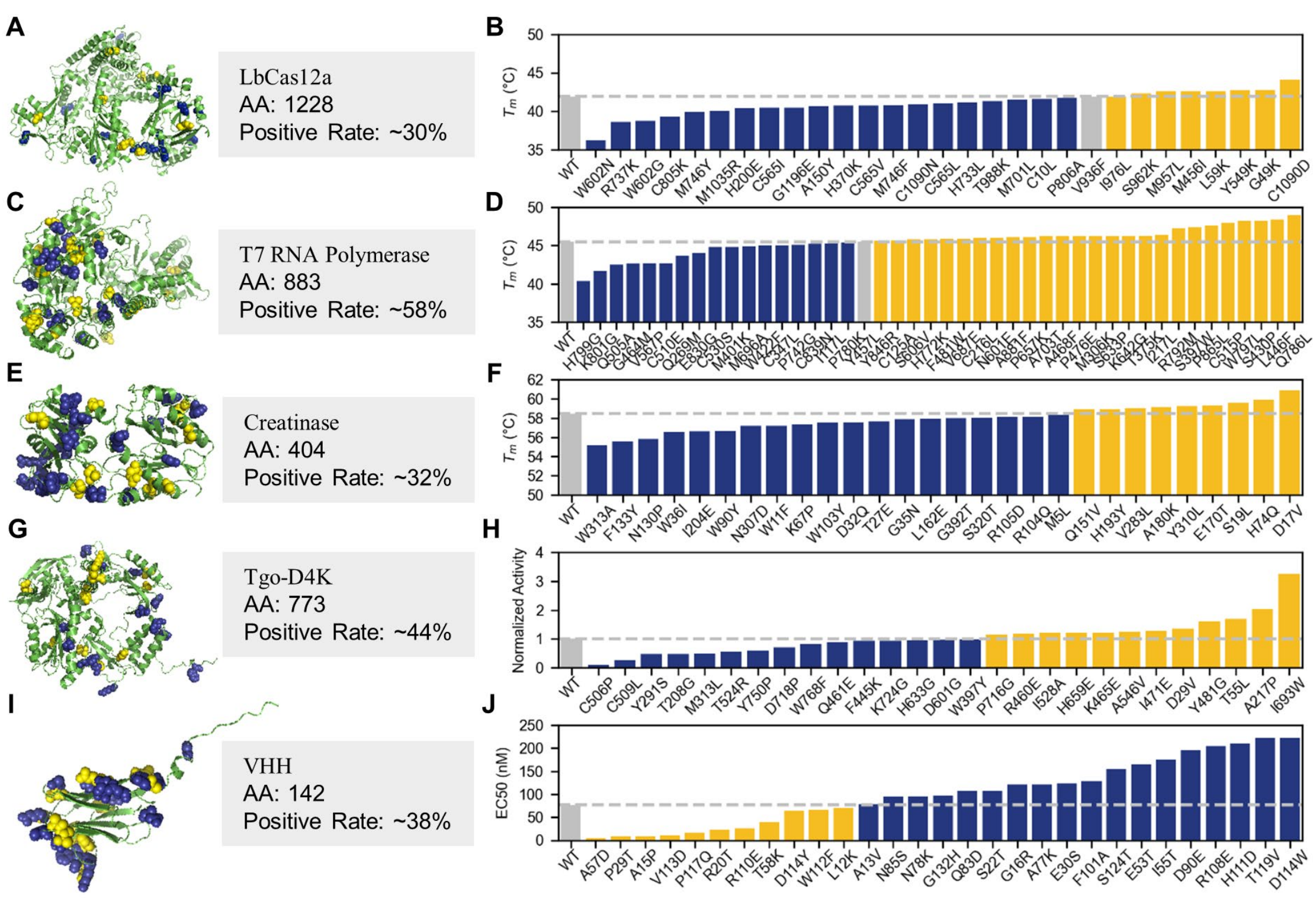

**Fig. 3**. **The structures and experimental results of single-site mutants predicted by PRIME for LbCas12a (A, B), T7 RNA polymerase (C, D), creatinase (E, F), non-natural nucleic acid polymerase (G, H) and VHH (I, J) are depicted.** The data points representing the mutations were systematically arranged in ascending order, with the corresponding value for the wild-type protein delineated by a gray bar for comparative purposes. Mutants that exhibited superior performance compared to their wild-type counterparts in terms of targeted attributes are highlighted in yellow, while negative mutants are shown in blue. The engineering goals varied between proteins for practical purposes: for LbCas12a, T7 RNA polymerase and creatinase, the objective was enhanced thermostability ($T_m$); for non-natural nucleic acid polymerase (Tgo-D4K), the aim was to accelerate the synthesis rate of 2'-fluoroarabino nucleic acid; and for VHH, the goal was to improve the tolerance ability under extreme alkaline pH conditions (EC50 of VHH binding to the antigen). All mutated structure were folded by Alphafold2. Detailed experimental data can be found in the separate Excel file in the supplementary materials.

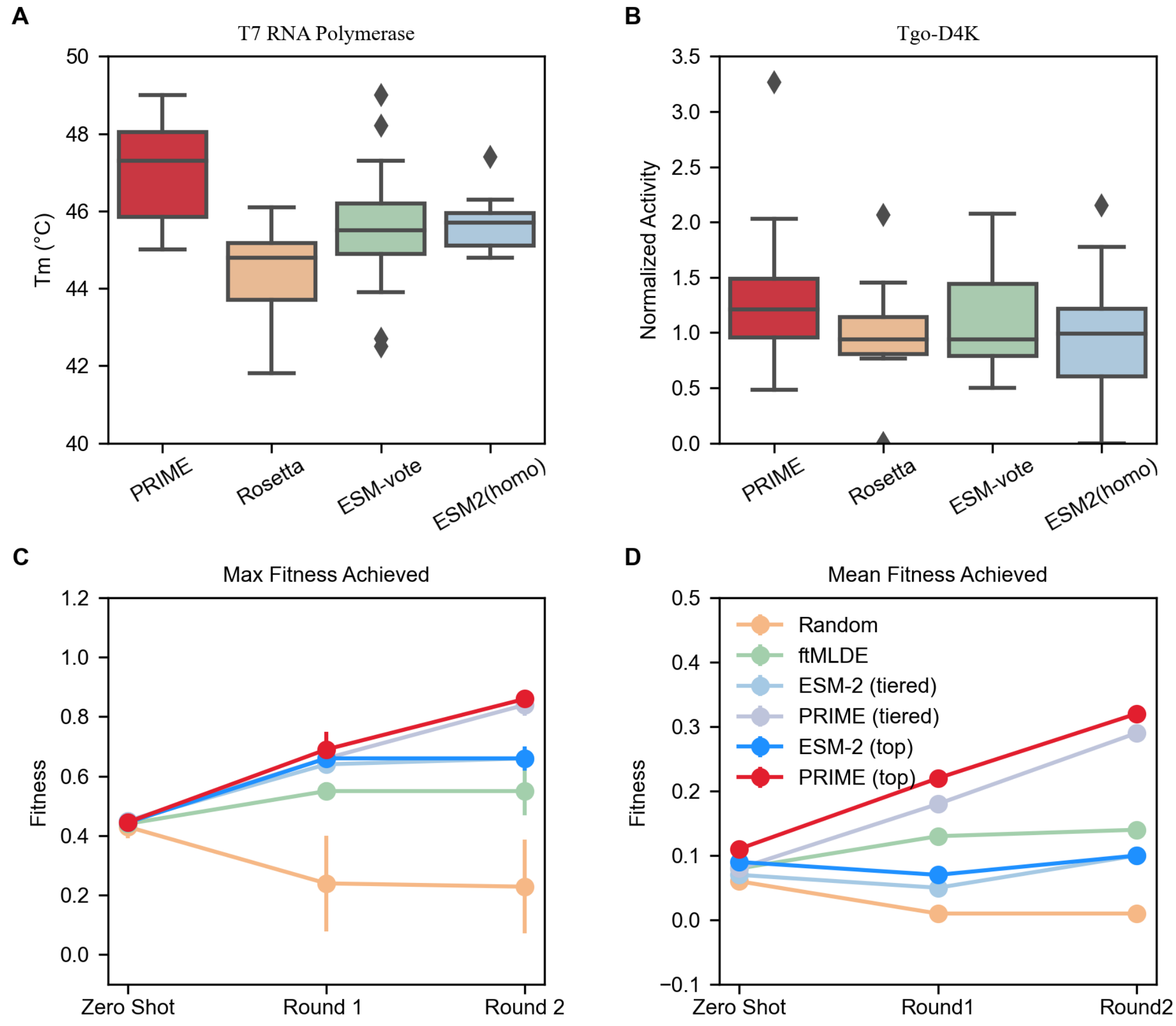

**Fig. 4**. **Comparative analysis of PRIME and different models through wet-lab experiments and in-silico benchmarking.** Panel A presents the comparative web-lab results for the top-15 single-point mutations in T7 RNA polymerase, and Panel B shows the results for Tgo-D4K, as determined by PRIME, Rosetta, ESM-vote, and ESM2(homo). Panels C and D depict the maximum (Panel C) and mean (Panel D) fitness outcomes obtained from in-silico directed evolution on the GB1 dataset, involving random mutagenesis, ftMLDE, ESM-2, and PRIME. For ESM-2 and PRIME, we examined both top-K sampling and the tiered sampling employed by ftMLDE(*77*).

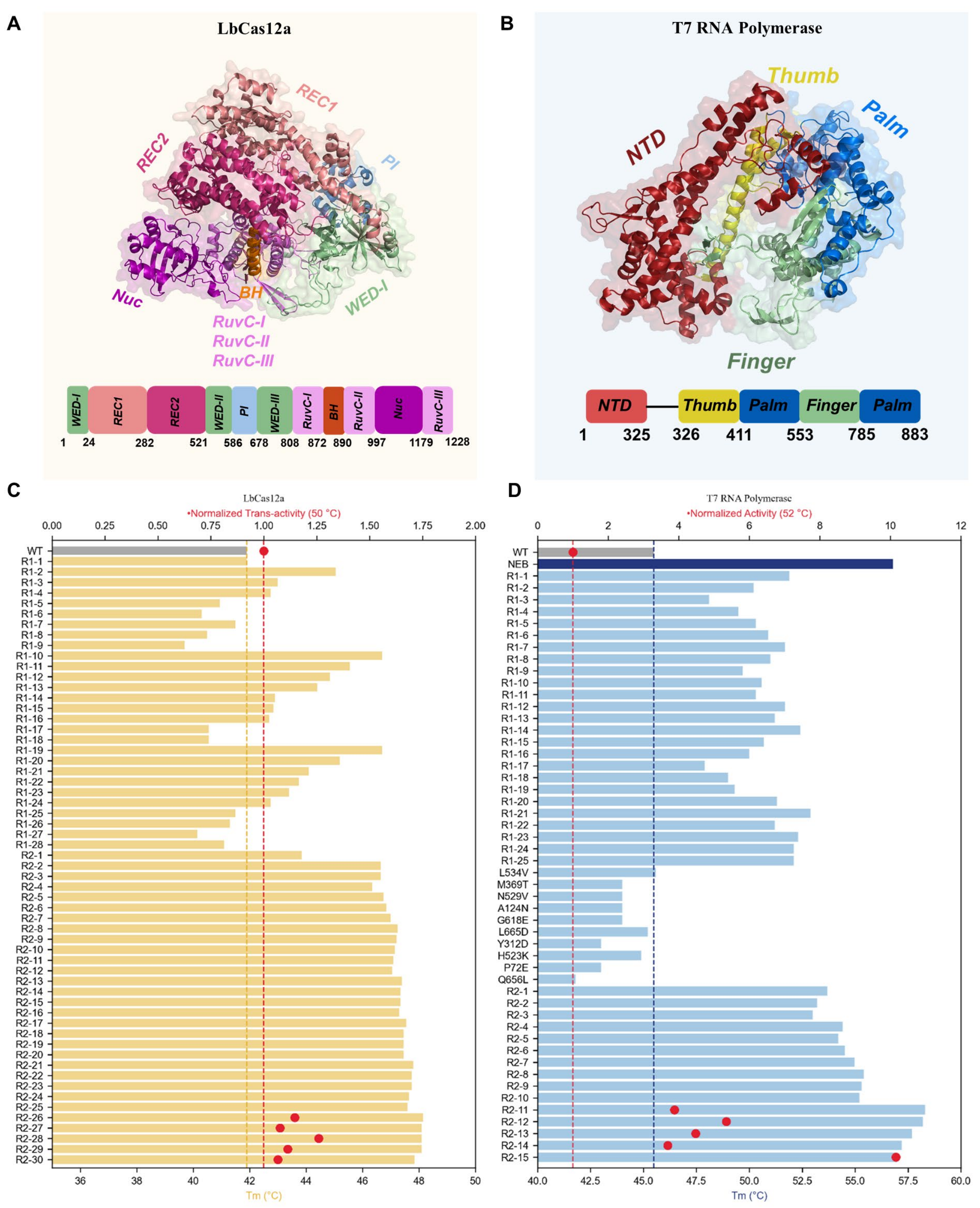

**Fig. 5. The protein structures and experimental results for multi-site mutants of LbCas12a (predicted by Alphafold2) (A and C) and T7 RNA polymerase (PDB ID: 1MSW) (B and D).** The functional domains of each protein are depicted beneath their structural diagrams. In the panels C and D, a red dashed line indicates the normalized activity level of the wild type. The activities of the five most thermostable mutants are marked with red dots to facilitate direct comparison with the wild type. The thermal stability ($T_m$) of the wild type is represented by a yellow dashed line for LbCas12a and a blue dashed line for T7 RNA polymerase. R1 and R2 represent the first and the second rounds of multipoint combination, respectively, with the numbers following them indicating the mutant indices.

**Acknowledgements**

This work was supported by the National Natural Science Foundation of China (Grant No. 12104295, 11974239 and 32471536), the National Key Research and Development Program of China (Grant No. 2021YFF1200200), the Innovation Program of Shanghai Municipal Education Commission (2019-01-07-00-02-E00076), Shanghai Jiao Tong University Scientific and Technological Innovation Funds (21X010200843), the Computational Biology Key Program of Shanghai Science and Technology Commission (23JS1400600), and Science and Technology Innovation Key R&D Program of Chongqing (CSTB2022TIAD-STX0017), the Student Innovation Center at Shanghai Jiao Tong University, and Shanghai Artificial Intelligence Laboratory. The engineering of VHH was supported by Changchun Genscience Pharmaceuticals Co., Ltd. We acknowledge Shanghai Artificial Intelligence Laboratory for computing resources.

**Author contributions**

Conceptualization: P.T., L.H.

Methodology: P.T., F.J, M.L.

Investigation: F.J., M.L., J.D., Y.Y, B.W., X.S., J.H., L.Z., Y.T., S.W., W.X., Y.P., J.S, L.K., Y.X., W.O., Z.H., G.F. Y.F., G.Y. J.L., Q.L.

Visualization: F.J., M.L.

Supervision: P.T., L.H., J.L., J.S.,

Writing—original draft: P.T., L.H.

Writing—review & editing: P.T., L.H., F.J, M.L.

**Competing interests**

The proteins mentioned in this article have both published patents and pending patent applications in China. These include LbCas12a, filed by ShanghaiTech University and Shanghai Jiao Tong

University (application number: 2023111716279); T7 RNA Polymerase and Creatinase, filed by Shanghai Matwings Technology Co., Ltd. (with both published and pending patents, publication numbers: CN117070493A, CN116694608A and CN117448306A); and VHH, filed by Changchun Genscience Pharmaceuticals Co., Ltd. The authors involved in these patent applications are Fan Jiang, Yuanxi Yu, Banghao Wu, Liqi Kang, Pan Tan, Jia Liu, and Liang Hong. The authors declare no other competing interests.

**Data and materials availability**

All data needed to evaluate the conclusions in this paper are present in the paper and/or the Supplementary Materials. Detailed information about the code used in this study is available at https://github.com/ai4protein/Pro-Prime.